\documentclass[aps,prl,twocolumn,showpacs,superscriptaddress,footinbib]{revtex4-1} 

\usepackage[english]{babel}
\usepackage{xcolor}
\usepackage{comment}
\usepackage{graphicx}
\usepackage{stackengine}
\usepackage{dcolumn}
\usepackage{bm}
\usepackage{bbm}
\usepackage[utf8]{inputenc}
\usepackage{amssymb}
\usepackage{amsmath}
\usepackage{bbold}
\usepackage{pstricks}
\usepackage{slashed}
\usepackage{braket}
\usepackage{hyperref}
\usepackage{natbib}
\usepackage{float}
\usepackage{verbatim}
\usepackage{siunitx}
\usepackage{lipsum}
\usepackage{slashed}
\usepackage{ulem}
\usepackage[left=1.85cm,right=1.85cm,top=1.85cm,bottom=1.85cm]{geometry}

\definecolor{mygreen}{rgb}{0,0.5,0}
\definecolor{myblue}{rgb}{0,0,0.75}
\definecolor{mymagenta}{cmyk}{0,1,0,0.12}
\definecolor{mygray}{rgb}{0.5,0.5,0.5}

\newcommand{\Fig}[1]{Fig.~\ref{#1}}

\begin{document}

\title{Error-corrected fermionic quantum processors with neutral atoms}

\author{Robert Ott}

\affiliation{Institute for Theoretical Physics, University of Innsbruck, Innsbruck, 6020, Austria}\affiliation{Institute for Quantum Optics and Quantum Information of the Austrian Academy of Sciences, Innsbruck, 6020, Austria}

\author{Daniel Gonz\'alez-Cuadra}
\affiliation{Institute for Theoretical Physics, University of Innsbruck, Innsbruck, 6020, Austria}
\affiliation{Institute for Quantum Optics and Quantum Information of the Austrian Academy of Sciences, Innsbruck, 6020, Austria}
\affiliation{Department of Physics, Harvard University, Cambridge, Massachusetts 02138, USA}
\affiliation{Instituto de F\'isica Fundamental, IFF-CSIC, Calle Serrano 113b, 28006 Madrid, Spain}

\author{Torsten V. Zache}
\affiliation{Institute for Theoretical Physics, University of Innsbruck, Innsbruck, 6020, Austria}
\affiliation{Institute for Quantum Optics and Quantum Information of the Austrian Academy of Sciences, Innsbruck, 6020, Austria}

\author{Peter Zoller}
\affiliation{Institute for Theoretical Physics, University of Innsbruck, Innsbruck, 6020, Austria}\affiliation{Institute for Quantum Optics and Quantum Information of the Austrian Academy of Sciences, Innsbruck, 6020, Austria}

\author{Adam M. Kaufman}
\affiliation{JILA, University of Colorado and National Institute of Standards and Technology,
and Department of Physics, University of Colorado, Boulder, Colorado 80309, USA}

\author{Hannes Pichler}
\email{hannes.pichler@uibk.ac.at}
\affiliation{Institute for Theoretical Physics, University of Innsbruck, Innsbruck, 6020, Austria}\affiliation{Institute for Quantum Optics and Quantum Information of the Austrian Academy of Sciences, Innsbruck, 6020, Austria}

\begin{abstract}
Many-body fermionic systems can be simulated in a hardware-efficient manner using a fermionic quantum processor. Neutral atoms trapped in optical potentials can realize such processors, where non-local fermionic statistics are guaranteed at the hardware level. Implementing quantum error correction in this setup is however challenging, due to the atom-number superselection present in atomic systems, that is, the impossibility of creating coherent superpositions of different particle numbers. In this work, we overcome this constraint and present a blueprint for an error-corrected fermionic quantum processor that can be implemented using current experimental capabilities. To achieve this, we first consider an ancillary set of fermionic modes and design a \textit{fermionic reference}, which we then use to construct superpositions of different numbers of \textit{referenced fermions}. This allows us to build logical fermionic modes that can be error corrected using standard atomic operations. Here, we focus on phase errors, which we expect to be a dominant source of errors in neutral-atom quantum processors. We then construct logical fermionic gates, and show their implementation for the logical particle-number conserving processes relevant for quantum simulation. Finally, our protocol is illustrated with a minimal fermionic circuit, where it leads to a quadratic suppression of the logical error rate.

\end{abstract}

\maketitle

\textit{Introduction.}-- Quantum computation~\cite{nielsen2010quantum,preskill2018quantum} promises solutions to difficult problems across many disciplines of natural science, ranging from high-energy to condensed matter physics and quantum chemistry~\cite{McArdle_2020, Altman_2021, Daley_2022, DiMeglio_2024}. While conventional quantum computers operate with qubits, many of these problems are naturally formulated in terms of fermionic particles. Encoding the fermionic statistics with qubits, however, represents a major challenge, especially in the presence of long-range interactions~\cite{verstraete2005mapping,zohar2018eliminating,derby2021compact,bravyi2017tapering,bravyi2002fermionic,chen2023equivalence}. To address this challenge, there has been a growing interest in developing programmable fermionic quantum processors, designed to naturally encode fermionic exchange statistics into their hardware architecture~\cite{gonzalez2023fermionic,Gkritsis_2024,arguello2019analogue,zache2023fermion}. Among the most promising approaches is the use of fermionic neutral atoms trapped in optical lattices~\cite{jordens2008mott,mazurenko2017cold,hirthe2023magnetically,xu2023frustration,brown2019bad,koepsell2019imaging,koepsell2021microscopic}, and, more recently, in programmable tweezer arrays~\cite{murmann2015two,Yan_2022,Spar_2022,Becher_2020,serwane2011deterministic}. Here, the inherent  indistinguishability of the atoms provides a direct and efficient access to simulating complex fermionic systems.

A key open question in the development of fermionic quantum processors is their compatibility with quantum error correction, a crucial ingredient to scaling quantum processors in the presence of noise~\cite{shor1995scheme,gottesman1997stabilizer}. In conventional qubit codes, as realized in systems based on trapped ions~\cite{ryan2024high,da2024demonstration,postler2022demonstration,postler2024demonstration,egan2021fault}, superconducting circuits~\cite{acharya2024quantum,putterman2024hardware,sivak2023real} or Rydberg atom arrays~\cite{bluvstein2024logical,reichardt2024logical,bedalov2024fault}, the strategy is to encode logical quantum information into suitably entangled states of several physical qubits. While there is a straightforward formal extension of these ideas to fermionic processors, i.e. by mapping qubit states to mode occupations, physical implementations of this mapping are hindered by the fundamental atom number superselection rule in atomic experiments - that is, the conservation of the total atom number. Circumventing these limitations is an outstanding challenge and existing ideas require advanced experimental capabilities~\cite{iemini2015localized,iemini2016dissipative,lang2015topological,buhler2014majorana}, such as coherent coupling to thermodynamically large reservoirs of molecular Bose-Einstein condensates~\cite{diehl2008quantum,diehl2011topology,jiang2011majorana, schuckert2024fermionqubitfaulttolerantquantumcomputing,holland2001resonance}.

In this paper, we present a novel proposal for overcoming the atom number superselection rule in existing neutral atom setups with finite atom numbers, and use it to design a blueprint for an error-corrected fermionic quantum processor. Our proposal liberates the physical fermions from their particle number conservation constraint through a key innovation: the use of a \textit{fermionic reference}. This consists of an ancillary set of fermionic modes that can interchange particles with the system modes. We show that, with a careful design of this exchange process, the quantum states of the ancillary modes serve as a phase reference, allowing to create and probe superpositions involving different numbers of \textit{referenced} fermions. In this sense, the fermionic reference plays a role analogous to a laser that acts as a phase reference for manipulations of optical coherence \footnote{In fact, our construction generalizes a subtle discussion of optical coherence~\cite{molmer1997optical} to fermionic system}. Our construction thus opens up a wide range of quantum information applications to fermionic atomic systems, and we apply it here to realize fermionic quantum error correction using neutral-atom arrays.

\begin{figure}[t!]
\includegraphics[width=0.97\columnwidth]{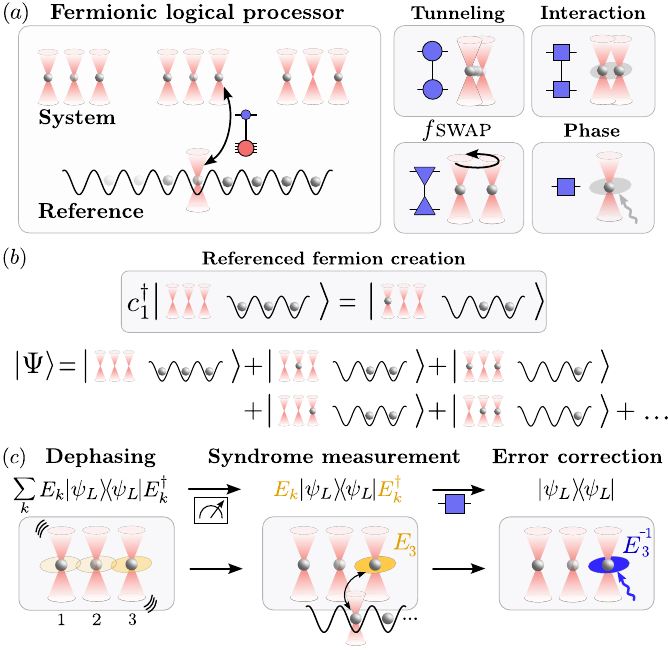}
\caption{\textit{Fermionic quantum processor.}~($a$) Fermionic processors based on neutral atoms in optical potentials realize physical operations such as inter-atomic interactions, single-atom phases, atom tunneling, and atom exchange [$f$\textsc{swap}]. ($b$)~We construct a fermionic reference, which allows to define referenced fermion operators  $c^{\small\dagger}$ and  $c$, and to realize processes that change their number without changing the physical atom number. This is required to create superpositions of states with different numbers of referenced fermions. ($c$)~Quantum error correction protects logical fermionic states against dephasing errors. Errors are identified using syndrome measurements (involving the reference) and are corrected with local operations.}
\label{fig:register}
\end{figure}

\textit{Fermionic neutral-atom arrays.}-- The envisioned setup based on spinless fermionic neutral atoms in optical potentials is outlined in \Fig{fig:register}. We use the ground state orbitals of optical tweezers to host fermionic modes~\cite{murmann2015two,Yan_2022, Spar_2022,Becher_2020,serwane2011deterministic} whose occupation with fermionic atoms defines the computational state~\cite{gonzalez2023fermionic}. Additionally, we represent the fermionic reference with a separate set of modes. These modes could be realized with a second array of tweezers, or represented by the lowest-band Wannier orbitals at different sites of an optical lattice~\cite{gross2017quantum}. We focus on the latter option to leverage the stability of this lattice. In this setup, we make use of the dynamical programmability of the system tweezers to transport and manipulate the atoms~\cite{bluvstein2022quantum,Kaufman_2021}, e.g., to implement tunneling operations by merging the tweezers~\cite{Kaufman_2014,murmann2015two,Becher_2020,Yan_2022,Spar_2022} or to implement interactions via Rydberg excitations~\cite{levine2019parallel,evered2023high,madjarov2020high,ma2023high,wilk2010entanglement,isenhower2010demonstration}. In addition, system and reference are interfaced with tunneling operations between tweezers and lattice sites~\cite{Young_2022,Tao_2024}, enabling exchange of particles between system and reference. In summary, these operations amount to the set of fermionic gates
$\mathcal{G}= \{ e^{i\theta n_i n_j}, e^{i\theta (f_i^\dagger f_j +\text{H.c.})}, e^{i\theta n_i} \}$
where $f^\dagger_i$($f_i$) denote the creation (annihilation) operators corresponding to a mode $i$ belonging either to a tweezer or lattice site, and $n_i = f^\dagger_if_i$. Throughout this paper, we assume that these operations can be efficiently implemented at the hardware level \cite{gonzalez2023fermionic}. Dominant errors arising in this setup are assumed to be phase errors from local fluctuations of the tweezer depth [Fig.~\ref{fig:register}(c)], while optical lattices are assumed to be robust, with leading errors given by homogeneous common-mode fluctuations. Focusing on such error models, we discuss how the use of a fermionic reference enables quantum error correction for neutral-atom fermionic processors.

\textit{Referenced fermion construction.}-- To be specific, we consider $M$ local fermionic modes with  fixed total \emph{physical} fermion number~$N$. We divide the $M$ modes into $M_{s}$ \textit{system} modes with annihilation operators $s_i$ ($i=1,.., M_s$) and $M_{\emph{r}}$ \textit{reference} modes with annihilation operators $r_i$ ($i=1,.., M_r$) \footnote{For simplicity we assume $M_r> N>M_s$, but discuss  generalizations in SM}. Our goal is to construct new fermionic modes which are not constrained by a particle number superselection rule~\footnote{We note that the entire construction here is employed for fermions, but can be applied to bosons as well.}.
To this end, we define reference ladder operators, $R$ and $R^\dagger$, via
\begin{align}\label{eq:R}
R &= \sum_{j=1}^{M_r}(1-\eta_{j+1})\,r_{j}\, \eta_{j-1}, 
\end{align}
where $\eta_i = r^\dagger_i r_i$ \footnote{We define the boundary term in \eqref{eq:R} via $\eta_{0}=1$, $\eta_{M_r+1}=0$.}. For each system fermion mode we define a corresponding \textit{referenced} fermion mode with creation and annihilation operators $c^\dagger_i = s_i^\dagger R$ and $c_i = R^\dagger s_i$, respectively. These operators move a physical fermion from the reference to the system (and vice versa), see \Fig{fig:register}($b$). Crucially, on the relevant Hilbert space  $\mathcal{H}$ (as defined below), they satisfy fermionic anti-commutation relations $\{c_i^\dagger,c_j\}\mathbf{P}= \mathbf{P}\{c_i^\dagger,c_j\}=\delta_{ij}\mathbf{P}$, which they inherit from the fermionic statistics of the physical atoms, and where $\mathbf{P}$ denotes the projector onto~$\mathcal{H}$. To define $\mathcal{H}$, we start by defining the state $\ket{\Omega} \equiv r_{N}^{\dagger}.. r_1^{\dagger}\ket{\mathrm{vac}}$, where $\ket{\mathrm{vac}}$ is the physical vacuum containing no atoms. We identify $\ket{\Omega}$ as the vacuum of the referenced fermions, since $c_i\ket{\Omega}=0$. The Hilbert space $\mathcal{H}$ is then spanned by the states that can be reached from $\ket{\Omega}$ by (multiple) applications of referenced fermion creation operators. It is straightforward to show that, on this space, $c_i$ and $c_{i}^\dag$ satisfy fermionic anti-communtation relation (see below/SM), giving $\mathcal{H}$ the structure of a fermionic Fock space for $N\geq M_s$.

The salient aspect of this definition is that all states in $\mathcal{H}$ have a particularly simple structure on the reference modes. To see this, note that the action of $R$ on $\ket{\Omega}$ simply removes the fermion from the mode with the largest index that is occupied, such that  $R^n\ket{\Omega}=r_{N-n}^\dag .. r^\dag_1 \ket{\rm{vac}}$. Similarly, on these states $R^\dag$ adds a fermion in the mode with smallest index that is unoccupied. As a consequence, all states containing $n$ referenced fermions in the system are associated with the same configuration of the reference modes, which contains a physical fermion in each reference mode with index $i\leq N-n$, and no fermion in reference modes with index $i>N-n$; see \Fig{fig:register}($b$)~\footnote{This particular set of reference states is related to the choice of $R$. It is perfectly possible to generalize this and choose different reference states and correspondingly $R$.}. Physically, this is reminiscent of a Fermi sea, where the particle in the mode with largest index is skimmed off by application of $R$. The structure of reference states implies the commutation relation $\mathbf{P}[R,R^\dagger]\mathbf{P} = 0$, 
which, in combination with the fermionic property $\{s_i,s_j^\dagger\}=\delta_{ij}$ of the original fermions, yields anti-commutation relations between referenced fermions. Importantly, while this construction of $\mathcal{H}$ allows only $N+1$ distinct reference configurations, the configuration of fermions in the system modes is not restricted, as illustrated in \Fig{fig:register}($b$).

An important feature of this construction is that number-conserving operations of referenced fermions do not involve the reference, and are thus equivalent to the corresponding operations in terms of system fermions, e.g., $c_i^\dagger c_j = s_i^\dagger s_j$. 
This enables all number-conserving operations in $\mathcal{G}$ at the level of referenced fermions, using the identical operations for the physical fermions. In addition, the reference allows to implement processes changing the number of referenced fermions, giving a fully universal gate-set for referenced fermions. These latter processes do not change the number of physical particles, but require acting on the reference modes. We give an explicit decomposition of such processes in terms of the gates in $\mathcal{G}$ next.
 
\textit{Realization of system-reference tunneling.}-- The main novel ingredient required for implementing our proposal is a physical operation that changes the number of referenced  particles. Here we focus on $D_i(\theta)\equiv\exp[i\theta(c^\dagger_i+c_i)]$, which is analogous to a Pauli-$X$ rotation in a qubit system. With this operation it is possible to implement our QEC scheme, but more general operations can also be obtained analogously. The unitary $D_i(\theta)$ amounts to a generalized system-reference tunneling involving the reference operators~$R$ and $R^\dagger$, therefore involving a single physical system mode as well as the entire reference. It can be realized by a sequence of tunneling operations, where a tweezer is coupled sequentially to each site of the reference in conjunction with density-interactions between lattice modes, e.g. realized with Rydberg interactions. As detailed in the Supplemental Material (SM) and shown in \Fig{fig:operations}($a$), when acting on $\mathcal{H}$ the unitary $D_i(\theta)$ can be decomposed as
\begin{align}
      e^{i\theta ( R^\dagger s_i +  s_i^\dagger R ) } &= \prod_{k=1}^{M_r}  \, e^{i\frac{\theta}{2}(s_i^\dagger r_k + \text{\text{H.c.}})}e^{i\pi \eta_k \eta_{k+1}}e^{i\pi \eta_k\eta_{k-1}}\nonumber\\
     &\times e^{-i\frac{\theta}{2}(s_i^\dagger r_k + \text{\text{H.c.}})}e^{i\pi \eta_k \eta_{k+1}}e^{i\pi \eta_k \eta_{k-1}} .
\end{align}
Here, we have made explicit use of the structure of reference states. The gate acts on all reference modes sequentially and therefore the number of two-mode gates is proportional to $M_r$.

\textit{Logical Fock space.}-- With the ability to create superpositions of fermion number sectors, we can translate the quantum error correction framework for qubits to our setup. To illustrate this, we focus on the simplest code, a repetition code, which is able to correct for local phase errors. We start by discussing this for a single logical fermionic mode. For this, we use three physical fermionic modes in conjunction with the reference. Specifically, the logical fermion annihilation operator is given by
\begin{align}
\label{eq:logical-operators}
    C&\equiv   i[ c_1c_2c_3 + c_1  c_2^\dagger c_3^\dagger+ c_1^\dagger c_2 c_3^\dagger+  c_1^\dagger c_2^\dagger c_3  ] \, ,
\end{align}
and $C^\dag$ is defined analogously. They act on the logical vacuum $\ket{0}_L \equiv \tfrac{1}{2}( 1 + ic_1^\dagger c_{2}^\dagger-ic_2^\dagger c_{3}^\dagger+ c_1^\dagger c_{3}^\dagger  )\ket{\Omega}$ as $C\ket{0}_L=0$, and $C^\dagger \ket{0}_L\equiv \ket{1}_L $ and fulfill anti-commutation relations~$\{C^\dagger,C\}=1$. The two logical fermionic states are stabilized by the operators
\begin{subequations}
\label{eq:stabilizer}
\begin{align}
    S_{12} &= \phantom{-}i (c_1+c_1^\dagger)\ (c_2+c_2^\dagger)\,,\\
    S_{23} &=  -i(c_2-c_2^\dagger)\ (c_3-c_3^\dagger)\, ,
\end{align}    
\end{subequations}
which commute with the logical fermion operators and define the local code space.

This construction can be generalized to multiple logical fermionic modes. For this, we partition the system modes into $M_L = M_\emph{s} / 3$ blocks of three system modes and label the logical modes with a block index $b=1,..,M_L$. Generalizing Eq.~\eqref{eq:logical-operators}, we define logical operators $C_b$ and $C_b^\dagger$, as well as the vacuum of $M_L$ logical modes $\ket{0_1,..,0_{M_L}}_L\!\! \sim\!\! \prod_{b} [ 1 + ic_{b,1}^\dagger c_{b,2}^\dagger-ic_{b,2}^\dagger c_{b,3}^\dagger+  c_{b,1}^\dagger c_{b,3}^\dagger]\ket{\Omega}$. From there,
we define a logical Fock space $\mathcal{H}^C$ spanned by the basis states 
\begin{align}
    \ket{n_{1},.., n_{M_L}}_L &= \big(C_{M_L}^\dagger\big)^{n_{M_L}}..\big(C_{1}^\dagger\big)^{n_{1}}\ket{0_1,..,0_{M_L}}_L ,
\end{align}
with $n_i\in \{0,1\}$, 
which are all stabilized by the operators $S_{12}^b$ and $S_{23}^b$ defined analogous to Eq.~\eqref{eq:stabilizer}. These logical states can be generated by a sequence of operations of the elementary gates or by using stabilizer measurements. Thus, we have constructed a fermionic code with fixed physical atom number with the help of a fermionic reference common to all logical modes~\footnote{Our construction can interpreted as encoding multiple logical modes in a single code block, evading the no-go theorem of Ref.~\cite{schuckert2024fermionqubitfaulttolerantquantumcomputing}.}.

\begin{figure}[t]
\includegraphics[width=1.\columnwidth]{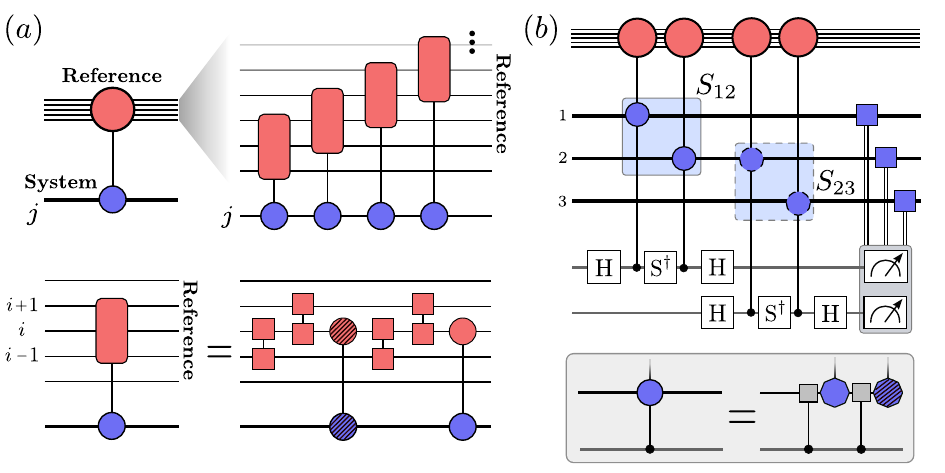}
\caption{\textit{Fermionic operations for error correction.}~($a$)~The fermionic reference enables operations that change the physical fermion number in the code. We show a decomposition of system-reference tunneling using $\mathcal{O}(M_r)$ elementary gates. ($b$)~To measure error syndromes, we employ system-reference tunneling and ancilla qubits. Measurement of the ancilla reveals the syndrome. The ancilla-controlled tunneling is decomposed in terms of density interactions and tunneling, see SM. Measuring both syndromes in each block allows to detect and correct single phase errors. Dashed gates represent the conjugate operation, circle/octagons/squares refer to tunneling and phases with $\theta=\tfrac{\pi}{2},\tfrac{\pi}{4},\pi$.}
\label{fig:operations}
\end{figure}

Logical operations that also do not change the number of logical fermions conserve the number of reference fermions, and therefore can be implemented without operating on the reference. Therefore, we restrict ourselves to Hilbert space sectors $\mathcal{H}^C_{N_L}$ with fixed logical particle number $N_L$ in the following. This is the relevant case for the quantum simulation of number-conserving interactions, as found in many physically relevant fermion models~\cite{McArdle_2020, Altman_2021, Daley_2022, DiMeglio_2024}. Furthermore, the fixed particle number sector is especially suited for our reference construction since, as we show below, this also enables correction of errors in the reference.

\textit{Quantum error correction.}-- The above construction forms an error-correcting code for the considered local phase errors, which amounts to the error set $\mathcal{E}=\{\mathbb{1},p_i\}$ with the local parity operators~$p_i=1-2s^\dagger_is_i$. It is easily checked that the Knill-Laflamme error correction condition~\cite{knill1997theory,nielsen2010quantum} is fulfilled for the set of errors $\mathcal{E}$ and the logical states, i.e. that $\bra{i_L}E^\dagger_{k}E_{l}\ket{j_L} \propto \delta_{ij}$ for any two errors~$E_k, E_l \in \mathcal{E}$. Therefore, phase errors are detectable and correctable. Specifically, $p_i$ flips a unique combination of stabilizer eigenvalues, e.g. $\{p_1,S_{12}\}=\{p_2,S_{12}\}=0$, $[p_3,S_{12}]=0$, such that the error $p_i$ can be uniquely inferred from the syndrome measurements.

We next detail the procedure for a single round of error correction as outlined in \Fig{fig:operations} with the help of an ancilla qubit \footnote{We assume that these ancilla qubits are error-free as they can in principle be encoded and error corrected using standard techniques~\cite{gottesman1997stabilizer}.}. To realize one round of error correction, we independently measure the two stabilizers of each block, as shown in \Fig{fig:operations}($c$) for the example of $S_{12}$. To achieve this, we map the stabilizer eigenvalues onto the state of the ancilla via the gate sequence $\mathrm{H}_a\times \mathrm{C}_a D_{2}(\tfrac{\pi}{2})\times\mathrm{S}_a^\dagger\times\mathrm{C}_a D_{1}(\tfrac{\pi}{2})\times\mathrm{H}_a$, where $\mathrm{H}_a$ is the Hadamard gate, $\mathrm{S}_a = \ket{0}\bra{0}_a +i \ket{1}\bra{1}_a$, and
\begin{align}
\label{eq:controlled-U}
\mathrm{C}_a D_{i}(\theta) = \ket{1}\bra{1}_a \otimes e^{i\theta\, (c_i + c_i^\dagger)} +\ket{0}\bra{0}_a \otimes \mathbb{1}\, ,
\end{align}
followed by a projective measurement of the ancilla qubit, indexed by the subscript~$a$. A decomposition of the ancilla-controlled tunneling~\ref{eq:controlled-U} in terms of system-reference tunneling and density interactions is shown in \Fig{fig:operations}($b$), see also SM for details, and a similar decomposition can be found to realize the second stabilizer $S_{23}$. The errors are subsequently corrected according to the measurement outcomes using local phase gates on the physical modes, as shown in \Fig{fig:operations}($c$). In our construction, a round of error correction can remove one phase error in each of the $M_L$ blocks with a total gate depth of~$\mathcal{O}(M_L)$~\footnote{Error correction for multiple blocks can be essentially parallelized.}.

\textit{Reference errors.}-- We now discuss the possibility of having errors also on the reference, where we distinguish two cases motivated by the physical properties of optical lattices. First, we consider global relative phases between reference and system, described by a unitary operator $\exp(i\epsilon \sum_j r^\dagger_j r_j)$, corresponding to global fluctuations of the lattice depth. This can be trivially accounted for, since it has the same effect as the previously considered phase errors on the system modes: due to the physical conservation of the atoms, this operators can be written as $\exp(-i\epsilon \sum_j s^\dagger_j s_j)$ (up to an irrelevant global phase), which for small $\epsilon \ll 1$ corresponds to the single-mode errors $p_i = 1- 2 s^\dagger _i s_i$ for all~$i$.

\begin{figure}[t]
\includegraphics[width=1.\columnwidth]{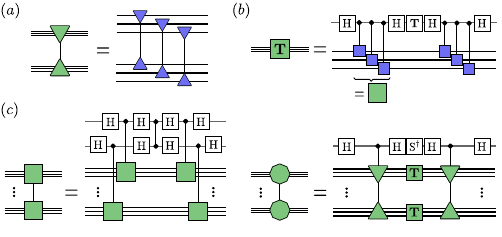}
\caption{\textit{Logical fermionic operations.}~($a$) Transversal implementation of the $f$\textsc{swap} gate, which encompasses the fermionic statistics. ($b$) The fermionic $\textbf{T}$-gate $\exp(i\tfrac{\pi}{4} N_i)$ can be realized with qubit $\textbf{T}$-gates on ancillas, see also Ref.~\cite{schuckert2024fermionqubitfaulttolerantquantumcomputing}. ($c$)~Similarly, two-mode density interaction $\exp(i\pi N_iN_j)$ and $\pi/4$-tunneling gates can be implemented with ancilla qubits using $\mathrm{CZ}$ and $\mathrm{S}$ gates. Ancillas are initialized in $\ket{0}$.}
\label{fig:gates}
\end{figure}

Going beyond the global reference errors, we discuss local errors in the reference next, $E_{Rj} = 1-2r^\dagger_j r_j$, for which our construction in $\mathcal{H}^C_{N_L}$ also fulfills the Knill-Laflamme error correction condition, see SM. By measuring the number of atoms $N_R=\sum_i r^\dagger_ir_i$ the reference state collapses into an eigenstate of $N_R$, and simultaneously removes the phase error, which acts only between different reference number states. While this collapses all superpositions between atom-number sectors on the system modes, the logical quantum information is preserved, but the state needs to be re-encoded at the end with $\mathcal{O}(M_L)$ operations.

\textit{Logical gate operations.}-- For logical computations on $\mathcal{H}^C_{N_L}$ we aim to construct the gate set
\begin{align}
    \mathcal{BK}_L'=\{e^{i\frac{\pi}{4}N_b},e^{i\pi N_bN_{b'}},e^{i\frac{\pi}{4}(C^\dagger_b C_{b'} + \text{\text{H.c.}})} \}\, ,
\end{align}
with logical (local) particle number $N_b=C^\dagger_b C_b$. Crucially, these logical operations can be implemented without involving the fermion reference, and hence without any additional overhead due to physical particle number conservation, see SM. 

In \Fig{fig:gates}, we show explicit decompositions of logical gates in terms of physical operations. Here, we allow the system to interact with a stable ancilla, on which gates can be performed with standard techniques~\cite{nielsen2010quantum}. The appeal of using a fermionic architecture lies in the efficient implementations of the fermionic exchange~$f$\textsc{swap}, which can be used to bring distant modes next to each other, including all fermionic exchange phases along their path~\cite{bravyi2002fermionic}. A key advantage of our hardware is that the~$f$\textsc{swap} operations can be implemented transversally, i.e. through the corresponding pair-wise operations on the same physical fermion modes of two different blocks, see \Fig{fig:gates}($a$). Here, this amounts to (classically) exchange all physical fermion modes of the two corresponding blocks, a process which can be highly parallelized using reconfigurable tweezer arrays~\cite{bluvstein2024logical}.

\begin{figure}[t]
\includegraphics[width=1.\columnwidth]{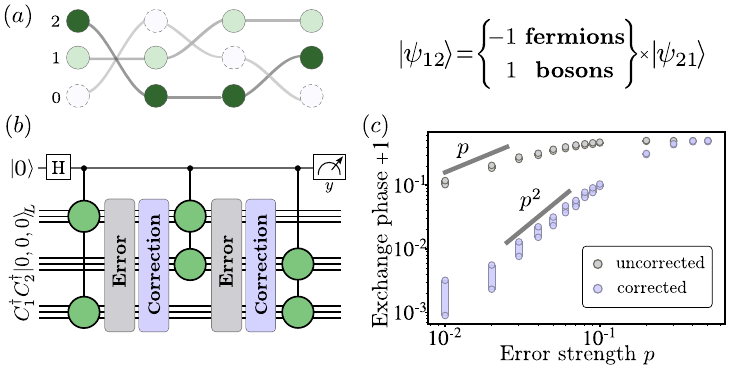}
\caption{\textit{Error-corrected fermionic circuit.} ($a$) We probe the logical fermion exchange with error-corrected computation. ($b$)~Controlled $\pi/2$-tunneling and ancilla measurements in the $y$-basis reveal the fermionic statistics. ($c$)~We simulate the circuit with layers of random phase errors (with single error probability $p$) and error correction. For small $p$, error correction decreases the logical error from $\mathcal{O}(p)$ (uncorrected) to $\mathcal{O}(p^2)$. Data shows $99\%$ (Clopper-Pearson) confidence intervals for $10^5$ realizations of the circuit. }
\label{fig:experiment}
\end{figure}

\textit{Minimal error-corrected fermionic quantum circuit.}-- To demonstrate fermionic quantum computation in conjunction with error correction, we propose a minimal experiment to test the fundamental fermionic statistics on the logical level. We propose to initialize three logical fermionic modes with total particle number $N_L$=$\,$2 and probe their fermionic statistics under particle exchange as illustrated in~\Fig{fig:experiment}($a$). The corresponding quantum circuit is shown in~\Fig{fig:experiment}($b$), where the required operations include tunneling operations between different logical fermionic modes. The tunneling operations are conditioned on the state of an ancilla such that the relative phase $\exp(i\Theta)$ between the two states $\ket{\psi_{12}} = C_2^\dagger C_1^\dagger\ket{0,0,0}_L$ and $\ket{\psi_{21}} = C_1^\dagger C_2^\dagger\ket{0,0,0}_L$ can be probed. 

To showcase such an experiment, we decompose the gates of the logical tunneling operations and simulate the resulting dynamics interspersed with several layers of phase errors and subsequent error correction as shown in~\Fig{fig:experiment}($b$). For every physical mode, errors are sampled such that in each error-layer $\exp(i\pi n_i)$ phases are applied with a given probability $p$. Single phase errors (per block) are corrected in the error-correcting layer while higher-weight errors result in logical errors, which propagate to the ancilla and corrupt the signal. \Fig{fig:experiment}($c$) shows the resulting exchange phase $\cos(i\Theta) = -\langle Y\rangle_a $ for various error strengths~$p$ and averaged over~$10^5$ realizations of the circuit. While the uncorrected result deviates from the ideal result $-1$ with $\mathcal{O}(p)$, error correction reduces this error to~$\mathcal{O}(p^2)$.

\textit{Conclusion \& Outlook.}-- Our work sets the stage for several directions of future investigations. While we demonstrated correction of phase errors with a repetition code, our reference construction allows us to straightforwardly import the entire qubit error-correction framework~\cite{gottesman1997stabilizer,nielsen2010quantum} for hardware-fermions, enabling generalization to more powerful codes  correcting more types of errors, including atom loss or leakage into other motional states of the system fermions \footnote{In the SM we illustrate this on the example of a fermionic Steane code.}, eventually pointing towards a fully fault-tolerant fermionic quantum processor. 

The key concept of our proposal is a fermionic reference to create superposition states with different fermion numbers. The crucial feature is the design of a set of reference states from which fermions can be extracted or added in a push/pop stack construction. We envision designing such a reference also for other experimental setups, beyond the concrete physical implementation proposed in this work. This includes alternative realizations of Fermi seas of neutral atoms in optical potentials~\cite{bayha2020observing}, as well as platforms with other fermionic particles, e.g. electrons in quantum dot arrays~\cite{hensgens2017quantum}.

Our work opens up the possibility for  quantitative comparisons of error corrected  fermionic- and conventional qubit-based quantum processors for simulating fermionic problems~\cite{verstraete2005mapping,zohar2018eliminating,derby2021compact,bravyi2017tapering,bravyi2002fermionic,chen2023equivalence,maskara2023programmable}. In this context, our work also motivates the development of novel quantum algorithms (and codes), which optimally leverage the intrinsic fermionic nature of the microscopic particles.

\textit{Note added:} In the final stages of completing this manuscript, we became aware of arXiv:2411.08955, which proposes QEC for fermionic processors by coupling to molecular BEC's as reservoirs for fermion pairs. In contrast, in our work the fermionic reference is a finite component of the processor that is interfaced algorithmically.

\textit{Acknowledgments.}-- We thank Alexander Schuckert for discussions.
Work in Innsbruck is funded by Horizon Europe programme HORIZON-CL4-2022-QUANTUM-02-SGA via the project 101113690 (PASQuanS2.1), the ERC Starting grant QARA (Grant No. 101041435), and by the Austrian Science
Fund (FWF) (Grant No. DOI 10.55776/COE1). D.G.-C. acknowledges support from the European Union’s Horizon Europe program under the Marie Skłodowska Curie Action PROGRAM (Grant No. 101150724). A. M. K. and H.P acknowledge support from the National Science Foundation QLCI (OMA-2016244). A. M. K. acknowledges support from the National Institute of Standards and Technology.

\bibliography{Bib}

\begin{thebibliography}{76}%
\makeatletter
\providecommand \@ifxundefined [1]{%
 \@ifx{#1\undefined}
}%
\providecommand \@ifnum [1]{%
 \ifnum #1\expandafter \@firstoftwo
 \else \expandafter \@secondoftwo
 \fi
}%
\providecommand \@ifx [1]{%
 \ifx #1\expandafter \@firstoftwo
 \else \expandafter \@secondoftwo
 \fi
}%
\providecommand \natexlab [1]{#1}%
\providecommand \enquote  [1]{``#1''}%
\providecommand \bibnamefont  [1]{#1}%
\providecommand \bibfnamefont [1]{#1}%
\providecommand \citenamefont [1]{#1}%
\providecommand \href@noop [0]{\@secondoftwo}%
\providecommand \href [0]{\begingroup \@sanitize@url \@href}%
\providecommand \@href[1]{\@@startlink{#1}\@@href}%
\providecommand \@@href[1]{\endgroup#1\@@endlink}%
\providecommand \@sanitize@url [0]{\catcode `\\12\catcode `\$12\catcode
  `\&12\catcode `\#12\catcode `\^12\catcode `\_12\catcode `\%12\relax}%
\providecommand \@@startlink[1]{}%
\providecommand \@@endlink[0]{}%
\providecommand \url  [0]{\begingroup\@sanitize@url \@url }%
\providecommand \@url [1]{\endgroup\@href {#1}{\urlprefix }}%
\providecommand \urlprefix  [0]{URL }%
\providecommand \Eprint [0]{\href }%
\providecommand \doibase [0]{http://dx.doi.org/}%
\providecommand \selectlanguage [0]{\@gobble}%
\providecommand \bibinfo  [0]{\@secondoftwo}%
\providecommand \bibfield  [0]{\@secondoftwo}%
\providecommand \translation [1]{[#1]}%
\providecommand \BibitemOpen [0]{}%
\providecommand \bibitemStop [0]{}%
\providecommand \bibitemNoStop [0]{.\EOS\space}%
\providecommand \EOS [0]{\spacefactor3000\relax}%
\providecommand \BibitemShut  [1]{\csname bibitem#1\endcsname}%
\let\auto@bib@innerbib\@empty
\bibitem [{\citenamefont {Nielsen}\ and\ \citenamefont
  {Chuang}(2010)}]{nielsen2010quantum}%
  \BibitemOpen
  \bibfield  {author} {\bibinfo {author} {\bibfnamefont {M.~A.}\ \bibnamefont
  {Nielsen}}\ and\ \bibinfo {author} {\bibfnamefont {I.~L.}\ \bibnamefont
  {Chuang}},\ }\href@noop {} {\emph {\bibinfo {title} {Quantum computation and
  quantum information}}}\ (\bibinfo  {publisher} {Cambridge university press},\
  \bibinfo {year} {2010})\BibitemShut {NoStop}%
\bibitem [{\citenamefont {Preskill}(2018)}]{preskill2018quantum}%
  \BibitemOpen
  \bibfield  {author} {\bibinfo {author} {\bibfnamefont {J.}~\bibnamefont
  {Preskill}},\ }\href@noop {} {\bibfield  {journal} {\bibinfo  {journal}
  {Quantum}\ }\textbf {\bibinfo {volume} {2}},\ \bibinfo {pages} {79} (\bibinfo
  {year} {2018})}\BibitemShut {NoStop}%
\bibitem [{\citenamefont {McArdle}\ \emph {et~al.}(2020)\citenamefont
  {McArdle}, \citenamefont {Endo}, \citenamefont {Aspuru-Guzik}, \citenamefont
  {Benjamin},\ and\ \citenamefont {Yuan}}]{McArdle_2020}%
  \BibitemOpen
  \bibfield  {author} {\bibinfo {author} {\bibfnamefont {S.}~\bibnamefont
  {McArdle}}, \bibinfo {author} {\bibfnamefont {S.}~\bibnamefont {Endo}},
  \bibinfo {author} {\bibfnamefont {A.}~\bibnamefont {Aspuru-Guzik}}, \bibinfo
  {author} {\bibfnamefont {S.~C.}\ \bibnamefont {Benjamin}}, \ and\ \bibinfo
  {author} {\bibfnamefont {X.}~\bibnamefont {Yuan}},\ }\href {\doibase
  10.1103/RevModPhys.92.015003} {\bibfield  {journal} {\bibinfo  {journal}
  {Rev. Mod. Phys.}\ }\textbf {\bibinfo {volume} {92}},\ \bibinfo {pages}
  {015003} (\bibinfo {year} {2020})}\BibitemShut {NoStop}%
\bibitem [{\citenamefont {Altman}\ \emph {et~al.}(2021)\citenamefont {Altman},
  \citenamefont {Brown}, \citenamefont {Carleo}, \citenamefont {Carr},
  \citenamefont {Demler}, \citenamefont {Chin}, \citenamefont {DeMarco},
  \citenamefont {Economou}, \citenamefont {Eriksson}, \citenamefont {Fu},
  \citenamefont {Greiner}, \citenamefont {Hazzard}, \citenamefont {Hulet},
  \citenamefont {Koll\'ar}, \citenamefont {Lev}, \citenamefont {Lukin},
  \citenamefont {Ma}, \citenamefont {Mi}, \citenamefont {Misra}, \citenamefont
  {Monroe}, \citenamefont {Murch}, \citenamefont {Nazario}, \citenamefont {Ni},
  \citenamefont {Potter}, \citenamefont {Roushan}, \citenamefont {Saffman},
  \citenamefont {Schleier-Smith}, \citenamefont {Siddiqi}, \citenamefont
  {Simmonds}, \citenamefont {Singh}, \citenamefont {Spielman}, \citenamefont
  {Temme}, \citenamefont {Weiss}, \citenamefont {Vu\ifmmode \check{c}\else
  \v{c}\fi{}kovi\ifmmode~\acute{c}\else \'{c}\fi{}}, \citenamefont
  {Vuleti\ifmmode~\acute{c}\else \'{c}\fi{}}, \citenamefont {Ye},\ and\
  \citenamefont {Zwierlein}}]{Altman_2021}%
  \BibitemOpen
  \bibfield  {author} {\bibinfo {author} {\bibfnamefont {E.}~\bibnamefont
  {Altman}}, \bibinfo {author} {\bibfnamefont {K.~R.}\ \bibnamefont {Brown}},
  \bibinfo {author} {\bibfnamefont {G.}~\bibnamefont {Carleo}}, \bibinfo
  {author} {\bibfnamefont {L.~D.}\ \bibnamefont {Carr}}, \bibinfo {author}
  {\bibfnamefont {E.}~\bibnamefont {Demler}}, \bibinfo {author} {\bibfnamefont
  {C.}~\bibnamefont {Chin}}, \bibinfo {author} {\bibfnamefont {B.}~\bibnamefont
  {DeMarco}}, \bibinfo {author} {\bibfnamefont {S.~E.}\ \bibnamefont
  {Economou}}, \bibinfo {author} {\bibfnamefont {M.~A.}\ \bibnamefont
  {Eriksson}}, \bibinfo {author} {\bibfnamefont {K.-M.~C.}\ \bibnamefont {Fu}},
  \bibinfo {author} {\bibfnamefont {M.}~\bibnamefont {Greiner}}, \bibinfo
  {author} {\bibfnamefont {K.~R.}\ \bibnamefont {Hazzard}}, \bibinfo {author}
  {\bibfnamefont {R.~G.}\ \bibnamefont {Hulet}}, \bibinfo {author}
  {\bibfnamefont {A.~J.}\ \bibnamefont {Koll\'ar}}, \bibinfo {author}
  {\bibfnamefont {B.~L.}\ \bibnamefont {Lev}}, \bibinfo {author} {\bibfnamefont
  {M.~D.}\ \bibnamefont {Lukin}}, \bibinfo {author} {\bibfnamefont
  {R.}~\bibnamefont {Ma}}, \bibinfo {author} {\bibfnamefont {X.}~\bibnamefont
  {Mi}}, \bibinfo {author} {\bibfnamefont {S.}~\bibnamefont {Misra}}, \bibinfo
  {author} {\bibfnamefont {C.}~\bibnamefont {Monroe}}, \bibinfo {author}
  {\bibfnamefont {K.}~\bibnamefont {Murch}}, \bibinfo {author} {\bibfnamefont
  {Z.}~\bibnamefont {Nazario}}, \bibinfo {author} {\bibfnamefont {K.-K.}\
  \bibnamefont {Ni}}, \bibinfo {author} {\bibfnamefont {A.~C.}\ \bibnamefont
  {Potter}}, \bibinfo {author} {\bibfnamefont {P.}~\bibnamefont {Roushan}},
  \bibinfo {author} {\bibfnamefont {M.}~\bibnamefont {Saffman}}, \bibinfo
  {author} {\bibfnamefont {M.}~\bibnamefont {Schleier-Smith}}, \bibinfo
  {author} {\bibfnamefont {I.}~\bibnamefont {Siddiqi}}, \bibinfo {author}
  {\bibfnamefont {R.}~\bibnamefont {Simmonds}}, \bibinfo {author}
  {\bibfnamefont {M.}~\bibnamefont {Singh}}, \bibinfo {author} {\bibfnamefont
  {I.}~\bibnamefont {Spielman}}, \bibinfo {author} {\bibfnamefont
  {K.}~\bibnamefont {Temme}}, \bibinfo {author} {\bibfnamefont {D.~S.}\
  \bibnamefont {Weiss}}, \bibinfo {author} {\bibfnamefont {J.}~\bibnamefont
  {Vu\ifmmode \check{c}\else \v{c}\fi{}kovi\ifmmode~\acute{c}\else
  \'{c}\fi{}}}, \bibinfo {author} {\bibfnamefont {V.}~\bibnamefont
  {Vuleti\ifmmode~\acute{c}\else \'{c}\fi{}}}, \bibinfo {author} {\bibfnamefont
  {J.}~\bibnamefont {Ye}}, \ and\ \bibinfo {author} {\bibfnamefont
  {M.}~\bibnamefont {Zwierlein}},\ }\href {\doibase
  10.1103/PRXQuantum.2.017003} {\bibfield  {journal} {\bibinfo  {journal} {PRX
  Quantum}\ }\textbf {\bibinfo {volume} {2}},\ \bibinfo {pages} {017003}
  (\bibinfo {year} {2021})}\BibitemShut {NoStop}%
\bibitem [{\citenamefont {Daley}\ \emph {et~al.}(2022)\citenamefont {Daley},
  \citenamefont {Bloch}, \citenamefont {Kokail}, \citenamefont {Flannigan},
  \citenamefont {Pearson}, \citenamefont {Troyer},\ and\ \citenamefont
  {Zoller}}]{Daley_2022}%
  \BibitemOpen
  \bibfield  {author} {\bibinfo {author} {\bibfnamefont {A.~J.}\ \bibnamefont
  {Daley}}, \bibinfo {author} {\bibfnamefont {I.}~\bibnamefont {Bloch}},
  \bibinfo {author} {\bibfnamefont {C.}~\bibnamefont {Kokail}}, \bibinfo
  {author} {\bibfnamefont {S.}~\bibnamefont {Flannigan}}, \bibinfo {author}
  {\bibfnamefont {N.}~\bibnamefont {Pearson}}, \bibinfo {author} {\bibfnamefont
  {M.}~\bibnamefont {Troyer}}, \ and\ \bibinfo {author} {\bibfnamefont
  {P.}~\bibnamefont {Zoller}},\ }\href {\doibase 10.1038/s41586-022-04940-6}
  {\bibfield  {journal} {\bibinfo  {journal} {Nature}\ }\textbf {\bibinfo
  {volume} {607}},\ \bibinfo {pages} {667} (\bibinfo {year}
  {2022})}\BibitemShut {NoStop}%
\bibitem [{\citenamefont {Di~Meglio}\ \emph {et~al.}(2024)\citenamefont
  {Di~Meglio}, \citenamefont {Jansen}, \citenamefont {Tavernelli},
  \citenamefont {Alexandrou}, \citenamefont {Arunachalam}, \citenamefont
  {Bauer}, \citenamefont {Borras}, \citenamefont {Carrazza}, \citenamefont
  {Crippa}, \citenamefont {Croft}, \citenamefont {de~Putter}, \citenamefont
  {Delgado}, \citenamefont {Dunjko}, \citenamefont {Egger}, \citenamefont
  {Fern\'andez-Combarro}, \citenamefont {Fuchs}, \citenamefont {Funcke},
  \citenamefont {Gonz\'alez-Cuadra}, \citenamefont {Grossi}, \citenamefont
  {Halimeh}, \citenamefont {Holmes}, \citenamefont {K\"uhn}, \citenamefont
  {Lacroix}, \citenamefont {Lewis}, \citenamefont {Lucchesi}, \citenamefont
  {Martinez}, \citenamefont {Meloni}, \citenamefont {Mezzacapo}, \citenamefont
  {Montangero}, \citenamefont {Nagano}, \citenamefont {Pascuzzi}, \citenamefont
  {Radescu}, \citenamefont {Ortega}, \citenamefont {Roggero}, \citenamefont
  {Schuhmacher}, \citenamefont {Seixas}, \citenamefont {Silvi}, \citenamefont
  {Spentzouris}, \citenamefont {Tacchino}, \citenamefont {Temme}, \citenamefont
  {Terashi}, \citenamefont {Tura}, \citenamefont {T\"uys\"uz}, \citenamefont
  {Vallecorsa}, \citenamefont {Wiese}, \citenamefont {Yoo},\ and\ \citenamefont
  {Zhang}}]{DiMeglio_2024}%
  \BibitemOpen
  \bibfield  {author} {\bibinfo {author} {\bibfnamefont {A.}~\bibnamefont
  {Di~Meglio}}, \bibinfo {author} {\bibfnamefont {K.}~\bibnamefont {Jansen}},
  \bibinfo {author} {\bibfnamefont {I.}~\bibnamefont {Tavernelli}}, \bibinfo
  {author} {\bibfnamefont {C.}~\bibnamefont {Alexandrou}}, \bibinfo {author}
  {\bibfnamefont {S.}~\bibnamefont {Arunachalam}}, \bibinfo {author}
  {\bibfnamefont {C.~W.}\ \bibnamefont {Bauer}}, \bibinfo {author}
  {\bibfnamefont {K.}~\bibnamefont {Borras}}, \bibinfo {author} {\bibfnamefont
  {S.}~\bibnamefont {Carrazza}}, \bibinfo {author} {\bibfnamefont
  {A.}~\bibnamefont {Crippa}}, \bibinfo {author} {\bibfnamefont
  {V.}~\bibnamefont {Croft}}, \bibinfo {author} {\bibfnamefont
  {R.}~\bibnamefont {de~Putter}}, \bibinfo {author} {\bibfnamefont
  {A.}~\bibnamefont {Delgado}}, \bibinfo {author} {\bibfnamefont
  {V.}~\bibnamefont {Dunjko}}, \bibinfo {author} {\bibfnamefont {D.~J.}\
  \bibnamefont {Egger}}, \bibinfo {author} {\bibfnamefont {E.}~\bibnamefont
  {Fern\'andez-Combarro}}, \bibinfo {author} {\bibfnamefont {E.}~\bibnamefont
  {Fuchs}}, \bibinfo {author} {\bibfnamefont {L.}~\bibnamefont {Funcke}},
  \bibinfo {author} {\bibfnamefont {D.}~\bibnamefont {Gonz\'alez-Cuadra}},
  \bibinfo {author} {\bibfnamefont {M.}~\bibnamefont {Grossi}}, \bibinfo
  {author} {\bibfnamefont {J.~C.}\ \bibnamefont {Halimeh}}, \bibinfo {author}
  {\bibfnamefont {Z.}~\bibnamefont {Holmes}}, \bibinfo {author} {\bibfnamefont
  {S.}~\bibnamefont {K\"uhn}}, \bibinfo {author} {\bibfnamefont
  {D.}~\bibnamefont {Lacroix}}, \bibinfo {author} {\bibfnamefont
  {R.}~\bibnamefont {Lewis}}, \bibinfo {author} {\bibfnamefont
  {D.}~\bibnamefont {Lucchesi}}, \bibinfo {author} {\bibfnamefont {M.~L.}\
  \bibnamefont {Martinez}}, \bibinfo {author} {\bibfnamefont {F.}~\bibnamefont
  {Meloni}}, \bibinfo {author} {\bibfnamefont {A.}~\bibnamefont {Mezzacapo}},
  \bibinfo {author} {\bibfnamefont {S.}~\bibnamefont {Montangero}}, \bibinfo
  {author} {\bibfnamefont {L.}~\bibnamefont {Nagano}}, \bibinfo {author}
  {\bibfnamefont {V.~R.}\ \bibnamefont {Pascuzzi}}, \bibinfo {author}
  {\bibfnamefont {V.}~\bibnamefont {Radescu}}, \bibinfo {author} {\bibfnamefont
  {E.~R.}\ \bibnamefont {Ortega}}, \bibinfo {author} {\bibfnamefont
  {A.}~\bibnamefont {Roggero}}, \bibinfo {author} {\bibfnamefont
  {J.}~\bibnamefont {Schuhmacher}}, \bibinfo {author} {\bibfnamefont
  {J.}~\bibnamefont {Seixas}}, \bibinfo {author} {\bibfnamefont
  {P.}~\bibnamefont {Silvi}}, \bibinfo {author} {\bibfnamefont
  {P.}~\bibnamefont {Spentzouris}}, \bibinfo {author} {\bibfnamefont
  {F.}~\bibnamefont {Tacchino}}, \bibinfo {author} {\bibfnamefont
  {K.}~\bibnamefont {Temme}}, \bibinfo {author} {\bibfnamefont
  {K.}~\bibnamefont {Terashi}}, \bibinfo {author} {\bibfnamefont
  {J.}~\bibnamefont {Tura}}, \bibinfo {author} {\bibfnamefont {C.}~\bibnamefont
  {T\"uys\"uz}}, \bibinfo {author} {\bibfnamefont {S.}~\bibnamefont
  {Vallecorsa}}, \bibinfo {author} {\bibfnamefont {U.-J.}\ \bibnamefont
  {Wiese}}, \bibinfo {author} {\bibfnamefont {S.}~\bibnamefont {Yoo}}, \ and\
  \bibinfo {author} {\bibfnamefont {J.}~\bibnamefont {Zhang}},\ }\href
  {\doibase 10.1103/PRXQuantum.5.037001} {\bibfield  {journal} {\bibinfo
  {journal} {PRX Quantum}\ }\textbf {\bibinfo {volume} {5}},\ \bibinfo {pages}
  {037001} (\bibinfo {year} {2024})}\BibitemShut {NoStop}%
\bibitem [{\citenamefont {Verstraete}\ and\ \citenamefont
  {Cirac}(2005)}]{verstraete2005mapping}%
  \BibitemOpen
  \bibfield  {author} {\bibinfo {author} {\bibfnamefont {F.}~\bibnamefont
  {Verstraete}}\ and\ \bibinfo {author} {\bibfnamefont {J.~I.}\ \bibnamefont
  {Cirac}},\ }\href@noop {} {\bibfield  {journal} {\bibinfo  {journal} {Journal
  of Statistical Mechanics: Theory and Experiment}\ }\textbf {\bibinfo {volume}
  {2005}},\ \bibinfo {pages} {P09012} (\bibinfo {year} {2005})}\BibitemShut
  {NoStop}%
\bibitem [{\citenamefont {Zohar}\ and\ \citenamefont
  {Cirac}(2018)}]{zohar2018eliminating}%
  \BibitemOpen
  \bibfield  {author} {\bibinfo {author} {\bibfnamefont {E.}~\bibnamefont
  {Zohar}}\ and\ \bibinfo {author} {\bibfnamefont {J.~I.}\ \bibnamefont
  {Cirac}},\ }\href@noop {} {\bibfield  {journal} {\bibinfo  {journal}
  {Physical Review B}\ }\textbf {\bibinfo {volume} {98}},\ \bibinfo {pages}
  {075119} (\bibinfo {year} {2018})}\BibitemShut {NoStop}%
\bibitem [{\citenamefont {Derby}\ \emph {et~al.}(2021)\citenamefont {Derby},
  \citenamefont {Klassen}, \citenamefont {Bausch},\ and\ \citenamefont
  {Cubitt}}]{derby2021compact}%
  \BibitemOpen
  \bibfield  {author} {\bibinfo {author} {\bibfnamefont {C.}~\bibnamefont
  {Derby}}, \bibinfo {author} {\bibfnamefont {J.}~\bibnamefont {Klassen}},
  \bibinfo {author} {\bibfnamefont {J.}~\bibnamefont {Bausch}}, \ and\ \bibinfo
  {author} {\bibfnamefont {T.}~\bibnamefont {Cubitt}},\ }\href@noop {}
  {\bibfield  {journal} {\bibinfo  {journal} {Physical Review B}\ }\textbf
  {\bibinfo {volume} {104}},\ \bibinfo {pages} {035118} (\bibinfo {year}
  {2021})}\BibitemShut {NoStop}%
\bibitem [{\citenamefont {Bravyi}\ \emph {et~al.}(2017)\citenamefont {Bravyi},
  \citenamefont {Gambetta}, \citenamefont {Mezzacapo},\ and\ \citenamefont
  {Temme}}]{bravyi2017tapering}%
  \BibitemOpen
  \bibfield  {author} {\bibinfo {author} {\bibfnamefont {S.}~\bibnamefont
  {Bravyi}}, \bibinfo {author} {\bibfnamefont {J.~M.}\ \bibnamefont
  {Gambetta}}, \bibinfo {author} {\bibfnamefont {A.}~\bibnamefont {Mezzacapo}},
  \ and\ \bibinfo {author} {\bibfnamefont {K.}~\bibnamefont {Temme}},\
  }\href@noop {} {\bibfield  {journal} {\bibinfo  {journal} {arXiv preprint
  arXiv:1701.08213}\ } (\bibinfo {year} {2017})}\BibitemShut {NoStop}%
\bibitem [{\citenamefont {Bravyi}\ and\ \citenamefont
  {Kitaev}(2002)}]{bravyi2002fermionic}%
  \BibitemOpen
  \bibfield  {author} {\bibinfo {author} {\bibfnamefont {S.~B.}\ \bibnamefont
  {Bravyi}}\ and\ \bibinfo {author} {\bibfnamefont {A.~Y.}\ \bibnamefont
  {Kitaev}},\ }\href@noop {} {\bibfield  {journal} {\bibinfo  {journal} {Annals
  of Physics}\ }\textbf {\bibinfo {volume} {298}},\ \bibinfo {pages} {210}
  (\bibinfo {year} {2002})}\BibitemShut {NoStop}%
\bibitem [{\citenamefont {Chen}\ and\ \citenamefont
  {Xu}(2023)}]{chen2023equivalence}%
  \BibitemOpen
  \bibfield  {author} {\bibinfo {author} {\bibfnamefont {Y.-A.}\ \bibnamefont
  {Chen}}\ and\ \bibinfo {author} {\bibfnamefont {Y.}~\bibnamefont {Xu}},\
  }\href@noop {} {\bibfield  {journal} {\bibinfo  {journal} {PRX Quantum}\
  }\textbf {\bibinfo {volume} {4}},\ \bibinfo {pages} {010326} (\bibinfo {year}
  {2023})}\BibitemShut {NoStop}%
\bibitem [{\citenamefont {González-Cuadra}\ \emph {et~al.}(2023)\citenamefont
  {González-Cuadra}, \citenamefont {Bluvstein}, \citenamefont {Kalinowski},
  \citenamefont {Kaubruegger}, \citenamefont {Maskara}, \citenamefont
  {Naldesi}, \citenamefont {Zache}, \citenamefont {Kaufman}, \citenamefont
  {Lukin}, \citenamefont {Pichler}, \citenamefont {Vermersch}, \citenamefont
  {Ye},\ and\ \citenamefont {Zoller}}]{gonzalez2023fermionic}%
  \BibitemOpen
  \bibfield  {author} {\bibinfo {author} {\bibfnamefont {D.}~\bibnamefont
  {González-Cuadra}}, \bibinfo {author} {\bibfnamefont {D.}~\bibnamefont
  {Bluvstein}}, \bibinfo {author} {\bibfnamefont {M.}~\bibnamefont
  {Kalinowski}}, \bibinfo {author} {\bibfnamefont {R.}~\bibnamefont
  {Kaubruegger}}, \bibinfo {author} {\bibfnamefont {N.}~\bibnamefont
  {Maskara}}, \bibinfo {author} {\bibfnamefont {P.}~\bibnamefont {Naldesi}},
  \bibinfo {author} {\bibfnamefont {T.~V.}\ \bibnamefont {Zache}}, \bibinfo
  {author} {\bibfnamefont {A.~M.}\ \bibnamefont {Kaufman}}, \bibinfo {author}
  {\bibfnamefont {M.~D.}\ \bibnamefont {Lukin}}, \bibinfo {author}
  {\bibfnamefont {H.}~\bibnamefont {Pichler}}, \bibinfo {author} {\bibfnamefont
  {B.}~\bibnamefont {Vermersch}}, \bibinfo {author} {\bibfnamefont
  {J.}~\bibnamefont {Ye}}, \ and\ \bibinfo {author} {\bibfnamefont
  {P.}~\bibnamefont {Zoller}},\ }\href {\doibase 10.1073/pnas.2304294120}
  {\bibfield  {journal} {\bibinfo  {journal} {Proceedings of the National
  Academy of Sciences}\ }\textbf {\bibinfo {volume} {120}},\ \bibinfo {pages}
  {e2304294120} (\bibinfo {year} {2023})}\BibitemShut {NoStop}%
\bibitem [{\citenamefont {Gkritsis}\ \emph {et~al.}(2024)\citenamefont
  {Gkritsis}, \citenamefont {Dux}, \citenamefont {Zhang}, \citenamefont {Jain},
  \citenamefont {Gogolin},\ and\ \citenamefont {Preiss}}]{Gkritsis_2024}%
  \BibitemOpen
  \bibfield  {author} {\bibinfo {author} {\bibfnamefont {F.}~\bibnamefont
  {Gkritsis}}, \bibinfo {author} {\bibfnamefont {D.}~\bibnamefont {Dux}},
  \bibinfo {author} {\bibfnamefont {J.}~\bibnamefont {Zhang}}, \bibinfo
  {author} {\bibfnamefont {N.}~\bibnamefont {Jain}}, \bibinfo {author}
  {\bibfnamefont {C.}~\bibnamefont {Gogolin}}, \ and\ \bibinfo {author}
  {\bibfnamefont {P.~M.}\ \bibnamefont {Preiss}},\ }\href@noop {} {} (\bibinfo
  {year} {2024}),\ \Eprint {http://arxiv.org/abs/2409.05663} {arXiv:2409.05663
  [cond-mat.quant-gas]} \BibitemShut {NoStop}%
\bibitem [{\citenamefont {Arg{\"u}ello-Luengo}\ \emph
  {et~al.}(2019)\citenamefont {Arg{\"u}ello-Luengo}, \citenamefont
  {Gonz{\'a}lez-Tudela}, \citenamefont {Shi}, \citenamefont {Zoller},\ and\
  \citenamefont {Cirac}}]{arguello2019analogue}%
  \BibitemOpen
  \bibfield  {author} {\bibinfo {author} {\bibfnamefont {J.}~\bibnamefont
  {Arg{\"u}ello-Luengo}}, \bibinfo {author} {\bibfnamefont {A.}~\bibnamefont
  {Gonz{\'a}lez-Tudela}}, \bibinfo {author} {\bibfnamefont {T.}~\bibnamefont
  {Shi}}, \bibinfo {author} {\bibfnamefont {P.}~\bibnamefont {Zoller}}, \ and\
  \bibinfo {author} {\bibfnamefont {J.~I.}\ \bibnamefont {Cirac}},\ }\href@noop
  {} {\bibfield  {journal} {\bibinfo  {journal} {Nature}\ }\textbf {\bibinfo
  {volume} {574}},\ \bibinfo {pages} {215} (\bibinfo {year}
  {2019})}\BibitemShut {NoStop}%
\bibitem [{\citenamefont {Zache}\ \emph {et~al.}(2023)\citenamefont {Zache},
  \citenamefont {Gonz{\'a}lez-Cuadra},\ and\ \citenamefont
  {Zoller}}]{zache2023fermion}%
  \BibitemOpen
  \bibfield  {author} {\bibinfo {author} {\bibfnamefont {T.~V.}\ \bibnamefont
  {Zache}}, \bibinfo {author} {\bibfnamefont {D.}~\bibnamefont
  {Gonz{\'a}lez-Cuadra}}, \ and\ \bibinfo {author} {\bibfnamefont
  {P.}~\bibnamefont {Zoller}},\ }\href@noop {} {\bibfield  {journal} {\bibinfo
  {journal} {Quantum}\ }\textbf {\bibinfo {volume} {7}},\ \bibinfo {pages}
  {1140} (\bibinfo {year} {2023})}\BibitemShut {NoStop}%
\bibitem [{\citenamefont {J{\"o}rdens}\ \emph {et~al.}(2008)\citenamefont
  {J{\"o}rdens}, \citenamefont {Strohmaier}, \citenamefont {G{\"u}nter},
  \citenamefont {Moritz},\ and\ \citenamefont {Esslinger}}]{jordens2008mott}%
  \BibitemOpen
  \bibfield  {author} {\bibinfo {author} {\bibfnamefont {R.}~\bibnamefont
  {J{\"o}rdens}}, \bibinfo {author} {\bibfnamefont {N.}~\bibnamefont
  {Strohmaier}}, \bibinfo {author} {\bibfnamefont {K.}~\bibnamefont
  {G{\"u}nter}}, \bibinfo {author} {\bibfnamefont {H.}~\bibnamefont {Moritz}},
  \ and\ \bibinfo {author} {\bibfnamefont {T.}~\bibnamefont {Esslinger}},\
  }\href@noop {} {\bibfield  {journal} {\bibinfo  {journal} {Nature}\ }\textbf
  {\bibinfo {volume} {455}},\ \bibinfo {pages} {204} (\bibinfo {year}
  {2008})}\BibitemShut {NoStop}%
\bibitem [{\citenamefont {Mazurenko}\ \emph {et~al.}(2017)\citenamefont
  {Mazurenko}, \citenamefont {Chiu}, \citenamefont {Ji}, \citenamefont
  {Parsons}, \citenamefont {Kan{\'a}sz-Nagy}, \citenamefont {Schmidt},
  \citenamefont {Grusdt}, \citenamefont {Demler}, \citenamefont {Greif},\ and\
  \citenamefont {Greiner}}]{mazurenko2017cold}%
  \BibitemOpen
  \bibfield  {author} {\bibinfo {author} {\bibfnamefont {A.}~\bibnamefont
  {Mazurenko}}, \bibinfo {author} {\bibfnamefont {C.~S.}\ \bibnamefont {Chiu}},
  \bibinfo {author} {\bibfnamefont {G.}~\bibnamefont {Ji}}, \bibinfo {author}
  {\bibfnamefont {M.~F.}\ \bibnamefont {Parsons}}, \bibinfo {author}
  {\bibfnamefont {M.}~\bibnamefont {Kan{\'a}sz-Nagy}}, \bibinfo {author}
  {\bibfnamefont {R.}~\bibnamefont {Schmidt}}, \bibinfo {author} {\bibfnamefont
  {F.}~\bibnamefont {Grusdt}}, \bibinfo {author} {\bibfnamefont
  {E.}~\bibnamefont {Demler}}, \bibinfo {author} {\bibfnamefont
  {D.}~\bibnamefont {Greif}}, \ and\ \bibinfo {author} {\bibfnamefont
  {M.}~\bibnamefont {Greiner}},\ }\href@noop {} {\bibfield  {journal} {\bibinfo
   {journal} {Nature}\ }\textbf {\bibinfo {volume} {545}},\ \bibinfo {pages}
  {462} (\bibinfo {year} {2017})}\BibitemShut {NoStop}%
\bibitem [{\citenamefont {Hirthe}\ \emph {et~al.}(2023)\citenamefont {Hirthe},
  \citenamefont {Chalopin}, \citenamefont {Bourgund}, \citenamefont
  {Bojovi{\'c}}, \citenamefont {Bohrdt}, \citenamefont {Demler}, \citenamefont
  {Grusdt}, \citenamefont {Bloch},\ and\ \citenamefont
  {Hilker}}]{hirthe2023magnetically}%
  \BibitemOpen
  \bibfield  {author} {\bibinfo {author} {\bibfnamefont {S.}~\bibnamefont
  {Hirthe}}, \bibinfo {author} {\bibfnamefont {T.}~\bibnamefont {Chalopin}},
  \bibinfo {author} {\bibfnamefont {D.}~\bibnamefont {Bourgund}}, \bibinfo
  {author} {\bibfnamefont {P.}~\bibnamefont {Bojovi{\'c}}}, \bibinfo {author}
  {\bibfnamefont {A.}~\bibnamefont {Bohrdt}}, \bibinfo {author} {\bibfnamefont
  {E.}~\bibnamefont {Demler}}, \bibinfo {author} {\bibfnamefont
  {F.}~\bibnamefont {Grusdt}}, \bibinfo {author} {\bibfnamefont
  {I.}~\bibnamefont {Bloch}}, \ and\ \bibinfo {author} {\bibfnamefont {T.~A.}\
  \bibnamefont {Hilker}},\ }\href@noop {} {\bibfield  {journal} {\bibinfo
  {journal} {Nature}\ }\textbf {\bibinfo {volume} {613}},\ \bibinfo {pages}
  {463} (\bibinfo {year} {2023})}\BibitemShut {NoStop}%
\bibitem [{\citenamefont {Xu}\ \emph {et~al.}(2023)\citenamefont {Xu},
  \citenamefont {Kendrick}, \citenamefont {Kale}, \citenamefont {Gang},
  \citenamefont {Ji}, \citenamefont {Scalettar}, \citenamefont {Lebrat},\ and\
  \citenamefont {Greiner}}]{xu2023frustration}%
  \BibitemOpen
  \bibfield  {author} {\bibinfo {author} {\bibfnamefont {M.}~\bibnamefont
  {Xu}}, \bibinfo {author} {\bibfnamefont {L.~H.}\ \bibnamefont {Kendrick}},
  \bibinfo {author} {\bibfnamefont {A.}~\bibnamefont {Kale}}, \bibinfo {author}
  {\bibfnamefont {Y.}~\bibnamefont {Gang}}, \bibinfo {author} {\bibfnamefont
  {G.}~\bibnamefont {Ji}}, \bibinfo {author} {\bibfnamefont {R.~T.}\
  \bibnamefont {Scalettar}}, \bibinfo {author} {\bibfnamefont {M.}~\bibnamefont
  {Lebrat}}, \ and\ \bibinfo {author} {\bibfnamefont {M.}~\bibnamefont
  {Greiner}},\ }\href@noop {} {\bibfield  {journal} {\bibinfo  {journal}
  {Nature}\ }\textbf {\bibinfo {volume} {620}},\ \bibinfo {pages} {971}
  (\bibinfo {year} {2023})}\BibitemShut {NoStop}%
\bibitem [{\citenamefont {Brown}\ \emph {et~al.}(2019)\citenamefont {Brown},
  \citenamefont {Mitra}, \citenamefont {Guardado-Sanchez}, \citenamefont
  {Nourafkan}, \citenamefont {Reymbaut}, \citenamefont {H{\'e}bert},
  \citenamefont {Bergeron}, \citenamefont {Tremblay}, \citenamefont {Kokalj},
  \citenamefont {Huse} \emph {et~al.}}]{brown2019bad}%
  \BibitemOpen
  \bibfield  {author} {\bibinfo {author} {\bibfnamefont {P.~T.}\ \bibnamefont
  {Brown}}, \bibinfo {author} {\bibfnamefont {D.}~\bibnamefont {Mitra}},
  \bibinfo {author} {\bibfnamefont {E.}~\bibnamefont {Guardado-Sanchez}},
  \bibinfo {author} {\bibfnamefont {R.}~\bibnamefont {Nourafkan}}, \bibinfo
  {author} {\bibfnamefont {A.}~\bibnamefont {Reymbaut}}, \bibinfo {author}
  {\bibfnamefont {C.-D.}\ \bibnamefont {H{\'e}bert}}, \bibinfo {author}
  {\bibfnamefont {S.}~\bibnamefont {Bergeron}}, \bibinfo {author}
  {\bibfnamefont {A.-M.}\ \bibnamefont {Tremblay}}, \bibinfo {author}
  {\bibfnamefont {J.}~\bibnamefont {Kokalj}}, \bibinfo {author} {\bibfnamefont
  {D.~A.}\ \bibnamefont {Huse}},  \emph {et~al.},\ }\href@noop {} {\bibfield
  {journal} {\bibinfo  {journal} {Science}\ }\textbf {\bibinfo {volume}
  {363}},\ \bibinfo {pages} {379} (\bibinfo {year} {2019})}\BibitemShut
  {NoStop}%
\bibitem [{\citenamefont {Koepsell}\ \emph {et~al.}(2019)\citenamefont
  {Koepsell}, \citenamefont {Vijayan}, \citenamefont {Sompet}, \citenamefont
  {Grusdt}, \citenamefont {Hilker}, \citenamefont {Demler}, \citenamefont
  {Salomon}, \citenamefont {Bloch},\ and\ \citenamefont
  {Gross}}]{koepsell2019imaging}%
  \BibitemOpen
  \bibfield  {author} {\bibinfo {author} {\bibfnamefont {J.}~\bibnamefont
  {Koepsell}}, \bibinfo {author} {\bibfnamefont {J.}~\bibnamefont {Vijayan}},
  \bibinfo {author} {\bibfnamefont {P.}~\bibnamefont {Sompet}}, \bibinfo
  {author} {\bibfnamefont {F.}~\bibnamefont {Grusdt}}, \bibinfo {author}
  {\bibfnamefont {T.~A.}\ \bibnamefont {Hilker}}, \bibinfo {author}
  {\bibfnamefont {E.}~\bibnamefont {Demler}}, \bibinfo {author} {\bibfnamefont
  {G.}~\bibnamefont {Salomon}}, \bibinfo {author} {\bibfnamefont
  {I.}~\bibnamefont {Bloch}}, \ and\ \bibinfo {author} {\bibfnamefont
  {C.}~\bibnamefont {Gross}},\ }\href@noop {} {\bibfield  {journal} {\bibinfo
  {journal} {Nature}\ }\textbf {\bibinfo {volume} {572}},\ \bibinfo {pages}
  {358} (\bibinfo {year} {2019})}\BibitemShut {NoStop}%
\bibitem [{\citenamefont {Koepsell}\ \emph {et~al.}(2021)\citenamefont
  {Koepsell}, \citenamefont {Bourgund}, \citenamefont {Sompet}, \citenamefont
  {Hirthe}, \citenamefont {Bohrdt}, \citenamefont {Wang}, \citenamefont
  {Grusdt}, \citenamefont {Demler}, \citenamefont {Salomon}, \citenamefont
  {Gross} \emph {et~al.}}]{koepsell2021microscopic}%
  \BibitemOpen
  \bibfield  {author} {\bibinfo {author} {\bibfnamefont {J.}~\bibnamefont
  {Koepsell}}, \bibinfo {author} {\bibfnamefont {D.}~\bibnamefont {Bourgund}},
  \bibinfo {author} {\bibfnamefont {P.}~\bibnamefont {Sompet}}, \bibinfo
  {author} {\bibfnamefont {S.}~\bibnamefont {Hirthe}}, \bibinfo {author}
  {\bibfnamefont {A.}~\bibnamefont {Bohrdt}}, \bibinfo {author} {\bibfnamefont
  {Y.}~\bibnamefont {Wang}}, \bibinfo {author} {\bibfnamefont {F.}~\bibnamefont
  {Grusdt}}, \bibinfo {author} {\bibfnamefont {E.}~\bibnamefont {Demler}},
  \bibinfo {author} {\bibfnamefont {G.}~\bibnamefont {Salomon}}, \bibinfo
  {author} {\bibfnamefont {C.}~\bibnamefont {Gross}},  \emph {et~al.},\
  }\href@noop {} {\bibfield  {journal} {\bibinfo  {journal} {Science}\ }\textbf
  {\bibinfo {volume} {374}},\ \bibinfo {pages} {82} (\bibinfo {year}
  {2021})}\BibitemShut {NoStop}%
\bibitem [{\citenamefont {Murmann}\ \emph {et~al.}(2015)\citenamefont
  {Murmann}, \citenamefont {Bergschneider}, \citenamefont {Klinkhamer},
  \citenamefont {Z\"urn}, \citenamefont {Lompe},\ and\ \citenamefont
  {Jochim}}]{murmann2015two}%
  \BibitemOpen
  \bibfield  {author} {\bibinfo {author} {\bibfnamefont {S.}~\bibnamefont
  {Murmann}}, \bibinfo {author} {\bibfnamefont {A.}~\bibnamefont
  {Bergschneider}}, \bibinfo {author} {\bibfnamefont {V.~M.}\ \bibnamefont
  {Klinkhamer}}, \bibinfo {author} {\bibfnamefont {G.}~\bibnamefont {Z\"urn}},
  \bibinfo {author} {\bibfnamefont {T.}~\bibnamefont {Lompe}}, \ and\ \bibinfo
  {author} {\bibfnamefont {S.}~\bibnamefont {Jochim}},\ }\href {\doibase
  10.1103/PhysRevLett.114.080402} {\bibfield  {journal} {\bibinfo  {journal}
  {Phys. Rev. Lett.}\ }\textbf {\bibinfo {volume} {114}},\ \bibinfo {pages}
  {080402} (\bibinfo {year} {2015})}\BibitemShut {NoStop}%
\bibitem [{\citenamefont {Yan}\ \emph {et~al.}(2022)\citenamefont {Yan},
  \citenamefont {Spar}, \citenamefont {Prichard}, \citenamefont {Chi},
  \citenamefont {Wei}, \citenamefont {Ibarra-Garc\'{\i}a-Padilla},
  \citenamefont {Hazzard},\ and\ \citenamefont {Bakr}}]{Yan_2022}%
  \BibitemOpen
  \bibfield  {author} {\bibinfo {author} {\bibfnamefont {Z.~Z.}\ \bibnamefont
  {Yan}}, \bibinfo {author} {\bibfnamefont {B.~M.}\ \bibnamefont {Spar}},
  \bibinfo {author} {\bibfnamefont {M.~L.}\ \bibnamefont {Prichard}}, \bibinfo
  {author} {\bibfnamefont {S.}~\bibnamefont {Chi}}, \bibinfo {author}
  {\bibfnamefont {H.-T.}\ \bibnamefont {Wei}}, \bibinfo {author} {\bibfnamefont
  {E.}~\bibnamefont {Ibarra-Garc\'{\i}a-Padilla}}, \bibinfo {author}
  {\bibfnamefont {K.~R.~A.}\ \bibnamefont {Hazzard}}, \ and\ \bibinfo {author}
  {\bibfnamefont {W.~S.}\ \bibnamefont {Bakr}},\ }\href {\doibase
  10.1103/PhysRevLett.129.123201} {\bibfield  {journal} {\bibinfo  {journal}
  {Phys. Rev. Lett.}\ }\textbf {\bibinfo {volume} {129}},\ \bibinfo {pages}
  {123201} (\bibinfo {year} {2022})}\BibitemShut {NoStop}%
\bibitem [{\citenamefont {Spar}\ \emph {et~al.}(2022)\citenamefont {Spar},
  \citenamefont {Guardado-Sanchez}, \citenamefont {Chi}, \citenamefont {Yan},\
  and\ \citenamefont {Bakr}}]{Spar_2022}%
  \BibitemOpen
  \bibfield  {author} {\bibinfo {author} {\bibfnamefont {B.~M.}\ \bibnamefont
  {Spar}}, \bibinfo {author} {\bibfnamefont {E.}~\bibnamefont
  {Guardado-Sanchez}}, \bibinfo {author} {\bibfnamefont {S.}~\bibnamefont
  {Chi}}, \bibinfo {author} {\bibfnamefont {Z.~Z.}\ \bibnamefont {Yan}}, \ and\
  \bibinfo {author} {\bibfnamefont {W.~S.}\ \bibnamefont {Bakr}},\ }\href
  {\doibase 10.1103/PhysRevLett.128.223202} {\bibfield  {journal} {\bibinfo
  {journal} {Phys. Rev. Lett.}\ }\textbf {\bibinfo {volume} {128}},\ \bibinfo
  {pages} {223202} (\bibinfo {year} {2022})}\BibitemShut {NoStop}%
\bibitem [{\citenamefont {Becher}\ \emph {et~al.}(2020)\citenamefont {Becher},
  \citenamefont {Sindici}, \citenamefont {Klemt}, \citenamefont {Jochim},
  \citenamefont {Daley},\ and\ \citenamefont {Preiss}}]{Becher_2020}%
  \BibitemOpen
  \bibfield  {author} {\bibinfo {author} {\bibfnamefont {J.~H.}\ \bibnamefont
  {Becher}}, \bibinfo {author} {\bibfnamefont {E.}~\bibnamefont {Sindici}},
  \bibinfo {author} {\bibfnamefont {R.}~\bibnamefont {Klemt}}, \bibinfo
  {author} {\bibfnamefont {S.}~\bibnamefont {Jochim}}, \bibinfo {author}
  {\bibfnamefont {A.~J.}\ \bibnamefont {Daley}}, \ and\ \bibinfo {author}
  {\bibfnamefont {P.~M.}\ \bibnamefont {Preiss}},\ }\href {\doibase
  10.1103/PhysRevLett.125.180402} {\bibfield  {journal} {\bibinfo  {journal}
  {Phys. Rev. Lett.}\ }\textbf {\bibinfo {volume} {125}},\ \bibinfo {pages}
  {180402} (\bibinfo {year} {2020})}\BibitemShut {NoStop}%
\bibitem [{\citenamefont {Serwane}\ \emph {et~al.}(2011)\citenamefont
  {Serwane}, \citenamefont {Z{\"u}rn}, \citenamefont {Lompe}, \citenamefont
  {Ottenstein}, \citenamefont {Wenz},\ and\ \citenamefont
  {Jochim}}]{serwane2011deterministic}%
  \BibitemOpen
  \bibfield  {author} {\bibinfo {author} {\bibfnamefont {F.}~\bibnamefont
  {Serwane}}, \bibinfo {author} {\bibfnamefont {G.}~\bibnamefont {Z{\"u}rn}},
  \bibinfo {author} {\bibfnamefont {T.}~\bibnamefont {Lompe}}, \bibinfo
  {author} {\bibfnamefont {T.}~\bibnamefont {Ottenstein}}, \bibinfo {author}
  {\bibfnamefont {A.}~\bibnamefont {Wenz}}, \ and\ \bibinfo {author}
  {\bibfnamefont {S.}~\bibnamefont {Jochim}},\ }\href@noop {} {\bibfield
  {journal} {\bibinfo  {journal} {Science}\ }\textbf {\bibinfo {volume}
  {332}},\ \bibinfo {pages} {336} (\bibinfo {year} {2011})}\BibitemShut
  {NoStop}%
\bibitem [{\citenamefont {Shor}(1995)}]{shor1995scheme}%
  \BibitemOpen
  \bibfield  {author} {\bibinfo {author} {\bibfnamefont {P.~W.}\ \bibnamefont
  {Shor}},\ }\href@noop {} {\bibfield  {journal} {\bibinfo  {journal} {Physical
  review A}\ }\textbf {\bibinfo {volume} {52}},\ \bibinfo {pages} {R2493}
  (\bibinfo {year} {1995})}\BibitemShut {NoStop}%
\bibitem [{\citenamefont {Gottesman}(1997)}]{gottesman1997stabilizer}%
  \BibitemOpen
  \bibfield  {author} {\bibinfo {author} {\bibfnamefont {D.}~\bibnamefont
  {Gottesman}},\ }\href@noop {} {\emph {\bibinfo {title} {Stabilizer codes and
  quantum error correction}}}\ (\bibinfo  {publisher} {California Institute of
  Technology},\ \bibinfo {year} {1997})\BibitemShut {NoStop}%
\bibitem [{\citenamefont {Ryan-Anderson}\ \emph {et~al.}(2024)\citenamefont
  {Ryan-Anderson}, \citenamefont {Brown}, \citenamefont {Baldwin},
  \citenamefont {Dreiling}, \citenamefont {Foltz}, \citenamefont {Gaebler},
  \citenamefont {Gatterman}, \citenamefont {Hewitt}, \citenamefont {Holliman},
  \citenamefont {Horst} \emph {et~al.}}]{ryan2024high}%
  \BibitemOpen
  \bibfield  {author} {\bibinfo {author} {\bibfnamefont {C.}~\bibnamefont
  {Ryan-Anderson}}, \bibinfo {author} {\bibfnamefont {N.}~\bibnamefont
  {Brown}}, \bibinfo {author} {\bibfnamefont {C.}~\bibnamefont {Baldwin}},
  \bibinfo {author} {\bibfnamefont {J.}~\bibnamefont {Dreiling}}, \bibinfo
  {author} {\bibfnamefont {C.}~\bibnamefont {Foltz}}, \bibinfo {author}
  {\bibfnamefont {J.}~\bibnamefont {Gaebler}}, \bibinfo {author} {\bibfnamefont
  {T.}~\bibnamefont {Gatterman}}, \bibinfo {author} {\bibfnamefont
  {N.}~\bibnamefont {Hewitt}}, \bibinfo {author} {\bibfnamefont
  {C.}~\bibnamefont {Holliman}}, \bibinfo {author} {\bibfnamefont
  {C.}~\bibnamefont {Horst}},  \emph {et~al.},\ }\href@noop {} {\bibfield
  {journal} {\bibinfo  {journal} {Science}\ }\textbf {\bibinfo {volume}
  {385}},\ \bibinfo {pages} {1327} (\bibinfo {year} {2024})}\BibitemShut
  {NoStop}%
\bibitem [{\citenamefont {Da~Silva}\ \emph {et~al.}(2024)\citenamefont
  {Da~Silva}, \citenamefont {Ryan-Anderson}, \citenamefont {Bello-Rivas},
  \citenamefont {Chernoguzov}, \citenamefont {Dreiling}, \citenamefont {Foltz},
  \citenamefont {Frachon}, \citenamefont {Gaebler}, \citenamefont {Gatterman},
  \citenamefont {Grans-Samuelsson} \emph {et~al.}}]{da2024demonstration}%
  \BibitemOpen
  \bibfield  {author} {\bibinfo {author} {\bibfnamefont {M.}~\bibnamefont
  {Da~Silva}}, \bibinfo {author} {\bibfnamefont {C.}~\bibnamefont
  {Ryan-Anderson}}, \bibinfo {author} {\bibfnamefont {J.}~\bibnamefont
  {Bello-Rivas}}, \bibinfo {author} {\bibfnamefont {A.}~\bibnamefont
  {Chernoguzov}}, \bibinfo {author} {\bibfnamefont {J.}~\bibnamefont
  {Dreiling}}, \bibinfo {author} {\bibfnamefont {C.}~\bibnamefont {Foltz}},
  \bibinfo {author} {\bibfnamefont {F.}~\bibnamefont {Frachon}}, \bibinfo
  {author} {\bibfnamefont {J.}~\bibnamefont {Gaebler}}, \bibinfo {author}
  {\bibfnamefont {T.}~\bibnamefont {Gatterman}}, \bibinfo {author}
  {\bibfnamefont {L.}~\bibnamefont {Grans-Samuelsson}},  \emph {et~al.},\
  }\href@noop {} {\bibfield  {journal} {\bibinfo  {journal} {arXiv preprint
  arXiv:2404.02280}\ } (\bibinfo {year} {2024})}\BibitemShut {NoStop}%
\bibitem [{\citenamefont {Postler}\ \emph {et~al.}(2022)\citenamefont
  {Postler}, \citenamefont {Heu$\beta$en}, \citenamefont {Pogorelov},
  \citenamefont {Rispler}, \citenamefont {Feldker}, \citenamefont {Meth},
  \citenamefont {Marciniak}, \citenamefont {Stricker}, \citenamefont
  {Ringbauer}, \citenamefont {Blatt} \emph
  {et~al.}}]{postler2022demonstration}%
  \BibitemOpen
  \bibfield  {author} {\bibinfo {author} {\bibfnamefont {L.}~\bibnamefont
  {Postler}}, \bibinfo {author} {\bibfnamefont {S.}~\bibnamefont
  {Heu$\beta$en}}, \bibinfo {author} {\bibfnamefont {I.}~\bibnamefont
  {Pogorelov}}, \bibinfo {author} {\bibfnamefont {M.}~\bibnamefont {Rispler}},
  \bibinfo {author} {\bibfnamefont {T.}~\bibnamefont {Feldker}}, \bibinfo
  {author} {\bibfnamefont {M.}~\bibnamefont {Meth}}, \bibinfo {author}
  {\bibfnamefont {C.~D.}\ \bibnamefont {Marciniak}}, \bibinfo {author}
  {\bibfnamefont {R.}~\bibnamefont {Stricker}}, \bibinfo {author}
  {\bibfnamefont {M.}~\bibnamefont {Ringbauer}}, \bibinfo {author}
  {\bibfnamefont {R.}~\bibnamefont {Blatt}},  \emph {et~al.},\ }\href@noop {}
  {\bibfield  {journal} {\bibinfo  {journal} {Nature}\ }\textbf {\bibinfo
  {volume} {605}},\ \bibinfo {pages} {675} (\bibinfo {year}
  {2022})}\BibitemShut {NoStop}%
\bibitem [{\citenamefont {Postler}\ \emph {et~al.}(2024)\citenamefont
  {Postler}, \citenamefont {Butt}, \citenamefont {Pogorelov}, \citenamefont
  {Marciniak}, \citenamefont {Heu{\ss}en}, \citenamefont {Blatt}, \citenamefont
  {Schindler}, \citenamefont {Rispler}, \citenamefont {M{\"u}ller},\ and\
  \citenamefont {Monz}}]{postler2024demonstration}%
  \BibitemOpen
  \bibfield  {author} {\bibinfo {author} {\bibfnamefont {L.}~\bibnamefont
  {Postler}}, \bibinfo {author} {\bibfnamefont {F.}~\bibnamefont {Butt}},
  \bibinfo {author} {\bibfnamefont {I.}~\bibnamefont {Pogorelov}}, \bibinfo
  {author} {\bibfnamefont {C.~D.}\ \bibnamefont {Marciniak}}, \bibinfo {author}
  {\bibfnamefont {S.}~\bibnamefont {Heu{\ss}en}}, \bibinfo {author}
  {\bibfnamefont {R.}~\bibnamefont {Blatt}}, \bibinfo {author} {\bibfnamefont
  {P.}~\bibnamefont {Schindler}}, \bibinfo {author} {\bibfnamefont
  {M.}~\bibnamefont {Rispler}}, \bibinfo {author} {\bibfnamefont
  {M.}~\bibnamefont {M{\"u}ller}}, \ and\ \bibinfo {author} {\bibfnamefont
  {T.}~\bibnamefont {Monz}},\ }\href@noop {} {\bibfield  {journal} {\bibinfo
  {journal} {PRX Quantum}\ }\textbf {\bibinfo {volume} {5}},\ \bibinfo {pages}
  {030326} (\bibinfo {year} {2024})}\BibitemShut {NoStop}%
\bibitem [{\citenamefont {Egan}\ \emph {et~al.}(2021)\citenamefont {Egan},
  \citenamefont {Debroy}, \citenamefont {Noel}, \citenamefont {Risinger},
  \citenamefont {Zhu}, \citenamefont {Biswas}, \citenamefont {Newman},
  \citenamefont {Li}, \citenamefont {Brown}, \citenamefont {Cetina} \emph
  {et~al.}}]{egan2021fault}%
  \BibitemOpen
  \bibfield  {author} {\bibinfo {author} {\bibfnamefont {L.}~\bibnamefont
  {Egan}}, \bibinfo {author} {\bibfnamefont {D.~M.}\ \bibnamefont {Debroy}},
  \bibinfo {author} {\bibfnamefont {C.}~\bibnamefont {Noel}}, \bibinfo {author}
  {\bibfnamefont {A.}~\bibnamefont {Risinger}}, \bibinfo {author}
  {\bibfnamefont {D.}~\bibnamefont {Zhu}}, \bibinfo {author} {\bibfnamefont
  {D.}~\bibnamefont {Biswas}}, \bibinfo {author} {\bibfnamefont
  {M.}~\bibnamefont {Newman}}, \bibinfo {author} {\bibfnamefont
  {M.}~\bibnamefont {Li}}, \bibinfo {author} {\bibfnamefont {K.~R.}\
  \bibnamefont {Brown}}, \bibinfo {author} {\bibfnamefont {M.}~\bibnamefont
  {Cetina}},  \emph {et~al.},\ }\href@noop {} {\bibfield  {journal} {\bibinfo
  {journal} {Nature}\ }\textbf {\bibinfo {volume} {598}},\ \bibinfo {pages}
  {281} (\bibinfo {year} {2021})}\BibitemShut {NoStop}%
\bibitem [{\citenamefont {Acharya}\ \emph {et~al.}(2024)\citenamefont
  {Acharya}, \citenamefont {Aghababaie-Beni}, \citenamefont {Aleiner},
  \citenamefont {Andersen}, \citenamefont {Ansmann}, \citenamefont {Arute},
  \citenamefont {Arya}, \citenamefont {Asfaw}, \citenamefont {Astrakhantsev},
  \citenamefont {Atalaya} \emph {et~al.}}]{acharya2024quantum}%
  \BibitemOpen
  \bibfield  {author} {\bibinfo {author} {\bibfnamefont {R.}~\bibnamefont
  {Acharya}}, \bibinfo {author} {\bibfnamefont {L.}~\bibnamefont
  {Aghababaie-Beni}}, \bibinfo {author} {\bibfnamefont {I.}~\bibnamefont
  {Aleiner}}, \bibinfo {author} {\bibfnamefont {T.~I.}\ \bibnamefont
  {Andersen}}, \bibinfo {author} {\bibfnamefont {M.}~\bibnamefont {Ansmann}},
  \bibinfo {author} {\bibfnamefont {F.}~\bibnamefont {Arute}}, \bibinfo
  {author} {\bibfnamefont {K.}~\bibnamefont {Arya}}, \bibinfo {author}
  {\bibfnamefont {A.}~\bibnamefont {Asfaw}}, \bibinfo {author} {\bibfnamefont
  {N.}~\bibnamefont {Astrakhantsev}}, \bibinfo {author} {\bibfnamefont
  {J.}~\bibnamefont {Atalaya}},  \emph {et~al.},\ }\href@noop {} {\bibfield
  {journal} {\bibinfo  {journal} {arXiv preprint arXiv:2408.13687}\ } (\bibinfo
  {year} {2024})}\BibitemShut {NoStop}%
\bibitem [{\citenamefont {Putterman}\ \emph {et~al.}(2024)\citenamefont
  {Putterman}, \citenamefont {Noh}, \citenamefont {Hann}, \citenamefont
  {MacCabe}, \citenamefont {Aghaeimeibodi}, \citenamefont {Patel},
  \citenamefont {Lee}, \citenamefont {Jones}, \citenamefont {Moradinejad},
  \citenamefont {Rodriguez} \emph {et~al.}}]{putterman2024hardware}%
  \BibitemOpen
  \bibfield  {author} {\bibinfo {author} {\bibfnamefont {H.}~\bibnamefont
  {Putterman}}, \bibinfo {author} {\bibfnamefont {K.}~\bibnamefont {Noh}},
  \bibinfo {author} {\bibfnamefont {C.~T.}\ \bibnamefont {Hann}}, \bibinfo
  {author} {\bibfnamefont {G.~S.}\ \bibnamefont {MacCabe}}, \bibinfo {author}
  {\bibfnamefont {S.}~\bibnamefont {Aghaeimeibodi}}, \bibinfo {author}
  {\bibfnamefont {R.~N.}\ \bibnamefont {Patel}}, \bibinfo {author}
  {\bibfnamefont {M.}~\bibnamefont {Lee}}, \bibinfo {author} {\bibfnamefont
  {W.~M.}\ \bibnamefont {Jones}}, \bibinfo {author} {\bibfnamefont
  {H.}~\bibnamefont {Moradinejad}}, \bibinfo {author} {\bibfnamefont
  {R.}~\bibnamefont {Rodriguez}},  \emph {et~al.},\ }\href@noop {} {\bibfield
  {journal} {\bibinfo  {journal} {arXiv preprint arXiv:2409.13025}\ } (\bibinfo
  {year} {2024})}\BibitemShut {NoStop}%
\bibitem [{\citenamefont {Sivak}\ \emph {et~al.}(2023)\citenamefont {Sivak},
  \citenamefont {Eickbusch}, \citenamefont {Royer}, \citenamefont {Singh},
  \citenamefont {Tsioutsios}, \citenamefont {Ganjam}, \citenamefont {Miano},
  \citenamefont {Brock}, \citenamefont {Ding}, \citenamefont {Frunzio} \emph
  {et~al.}}]{sivak2023real}%
  \BibitemOpen
  \bibfield  {author} {\bibinfo {author} {\bibfnamefont {V.}~\bibnamefont
  {Sivak}}, \bibinfo {author} {\bibfnamefont {A.}~\bibnamefont {Eickbusch}},
  \bibinfo {author} {\bibfnamefont {B.}~\bibnamefont {Royer}}, \bibinfo
  {author} {\bibfnamefont {S.}~\bibnamefont {Singh}}, \bibinfo {author}
  {\bibfnamefont {I.}~\bibnamefont {Tsioutsios}}, \bibinfo {author}
  {\bibfnamefont {S.}~\bibnamefont {Ganjam}}, \bibinfo {author} {\bibfnamefont
  {A.}~\bibnamefont {Miano}}, \bibinfo {author} {\bibfnamefont
  {B.}~\bibnamefont {Brock}}, \bibinfo {author} {\bibfnamefont
  {A.}~\bibnamefont {Ding}}, \bibinfo {author} {\bibfnamefont {L.}~\bibnamefont
  {Frunzio}},  \emph {et~al.},\ }\href@noop {} {\bibfield  {journal} {\bibinfo
  {journal} {Nature}\ }\textbf {\bibinfo {volume} {616}},\ \bibinfo {pages}
  {50} (\bibinfo {year} {2023})}\BibitemShut {NoStop}%
\bibitem [{\citenamefont {Bluvstein}\ \emph {et~al.}(2024)\citenamefont
  {Bluvstein}, \citenamefont {Evered}, \citenamefont {Geim}, \citenamefont
  {Li}, \citenamefont {Zhou}, \citenamefont {Manovitz}, \citenamefont {Ebadi},
  \citenamefont {Cain}, \citenamefont {Kalinowski}, \citenamefont {Hangleiter}
  \emph {et~al.}}]{bluvstein2024logical}%
  \BibitemOpen
  \bibfield  {author} {\bibinfo {author} {\bibfnamefont {D.}~\bibnamefont
  {Bluvstein}}, \bibinfo {author} {\bibfnamefont {S.~J.}\ \bibnamefont
  {Evered}}, \bibinfo {author} {\bibfnamefont {A.~A.}\ \bibnamefont {Geim}},
  \bibinfo {author} {\bibfnamefont {S.~H.}\ \bibnamefont {Li}}, \bibinfo
  {author} {\bibfnamefont {H.}~\bibnamefont {Zhou}}, \bibinfo {author}
  {\bibfnamefont {T.}~\bibnamefont {Manovitz}}, \bibinfo {author}
  {\bibfnamefont {S.}~\bibnamefont {Ebadi}}, \bibinfo {author} {\bibfnamefont
  {M.}~\bibnamefont {Cain}}, \bibinfo {author} {\bibfnamefont {M.}~\bibnamefont
  {Kalinowski}}, \bibinfo {author} {\bibfnamefont {D.}~\bibnamefont
  {Hangleiter}},  \emph {et~al.},\ }\href@noop {} {\bibfield  {journal}
  {\bibinfo  {journal} {Nature}\ }\textbf {\bibinfo {volume} {626}},\ \bibinfo
  {pages} {58} (\bibinfo {year} {2024})}\BibitemShut {NoStop}%
\bibitem [{\citenamefont {Reichardt}\ \emph {et~al.}(2024)\citenamefont
  {Reichardt}, \citenamefont {Paetznick}, \citenamefont {Aasen}, \citenamefont
  {Basov}, \citenamefont {Bello-Rivas}, \citenamefont {Bonderson},
  \citenamefont {Chao}, \citenamefont {van Dam}, \citenamefont {Hastings},
  \citenamefont {Paz} \emph {et~al.}}]{reichardt2024logical}%
  \BibitemOpen
  \bibfield  {author} {\bibinfo {author} {\bibfnamefont {B.~W.}\ \bibnamefont
  {Reichardt}}, \bibinfo {author} {\bibfnamefont {A.}~\bibnamefont
  {Paetznick}}, \bibinfo {author} {\bibfnamefont {D.}~\bibnamefont {Aasen}},
  \bibinfo {author} {\bibfnamefont {I.}~\bibnamefont {Basov}}, \bibinfo
  {author} {\bibfnamefont {J.~M.}\ \bibnamefont {Bello-Rivas}}, \bibinfo
  {author} {\bibfnamefont {P.}~\bibnamefont {Bonderson}}, \bibinfo {author}
  {\bibfnamefont {R.}~\bibnamefont {Chao}}, \bibinfo {author} {\bibfnamefont
  {W.}~\bibnamefont {van Dam}}, \bibinfo {author} {\bibfnamefont {M.~B.}\
  \bibnamefont {Hastings}}, \bibinfo {author} {\bibfnamefont {A.}~\bibnamefont
  {Paz}},  \emph {et~al.},\ }\href@noop {} {\bibfield  {journal} {\bibinfo
  {journal} {arXiv preprint arXiv:2411.11822}\ } (\bibinfo {year}
  {2024})}\BibitemShut {NoStop}%
\bibitem [{\citenamefont {Bedalov}\ \emph {et~al.}(2024)\citenamefont
  {Bedalov}, \citenamefont {Blakely}, \citenamefont {Buttler}, \citenamefont
  {Carnahan}, \citenamefont {Chong}, \citenamefont {Chung}, \citenamefont
  {Cole}, \citenamefont {Goiporia}, \citenamefont {Gokhale}, \citenamefont
  {Heim} \emph {et~al.}}]{bedalov2024fault}%
  \BibitemOpen
  \bibfield  {author} {\bibinfo {author} {\bibfnamefont {M.}~\bibnamefont
  {Bedalov}}, \bibinfo {author} {\bibfnamefont {M.}~\bibnamefont {Blakely}},
  \bibinfo {author} {\bibfnamefont {P.}~\bibnamefont {Buttler}}, \bibinfo
  {author} {\bibfnamefont {C.}~\bibnamefont {Carnahan}}, \bibinfo {author}
  {\bibfnamefont {F.~T.}\ \bibnamefont {Chong}}, \bibinfo {author}
  {\bibfnamefont {W.~C.}\ \bibnamefont {Chung}}, \bibinfo {author}
  {\bibfnamefont {D.~C.}\ \bibnamefont {Cole}}, \bibinfo {author}
  {\bibfnamefont {P.}~\bibnamefont {Goiporia}}, \bibinfo {author}
  {\bibfnamefont {P.}~\bibnamefont {Gokhale}}, \bibinfo {author} {\bibfnamefont
  {B.}~\bibnamefont {Heim}},  \emph {et~al.},\ }\href@noop {} {\bibfield
  {journal} {\bibinfo  {journal} {arXiv preprint arXiv:2412.07670}\ } (\bibinfo
  {year} {2024})}\BibitemShut {NoStop}%
\bibitem [{\citenamefont {Iemini}\ \emph {et~al.}(2015)\citenamefont {Iemini},
  \citenamefont {Mazza}, \citenamefont {Rossini}, \citenamefont {Fazio},\ and\
  \citenamefont {Diehl}}]{iemini2015localized}%
  \BibitemOpen
  \bibfield  {author} {\bibinfo {author} {\bibfnamefont {F.}~\bibnamefont
  {Iemini}}, \bibinfo {author} {\bibfnamefont {L.}~\bibnamefont {Mazza}},
  \bibinfo {author} {\bibfnamefont {D.}~\bibnamefont {Rossini}}, \bibinfo
  {author} {\bibfnamefont {R.}~\bibnamefont {Fazio}}, \ and\ \bibinfo {author}
  {\bibfnamefont {S.}~\bibnamefont {Diehl}},\ }\href@noop {} {\bibfield
  {journal} {\bibinfo  {journal} {Physical review letters}\ }\textbf {\bibinfo
  {volume} {115}},\ \bibinfo {pages} {156402} (\bibinfo {year}
  {2015})}\BibitemShut {NoStop}%
\bibitem [{\citenamefont {Iemini}\ \emph {et~al.}(2016)\citenamefont {Iemini},
  \citenamefont {Rossini}, \citenamefont {Fazio}, \citenamefont {Diehl},\ and\
  \citenamefont {Mazza}}]{iemini2016dissipative}%
  \BibitemOpen
  \bibfield  {author} {\bibinfo {author} {\bibfnamefont {F.}~\bibnamefont
  {Iemini}}, \bibinfo {author} {\bibfnamefont {D.}~\bibnamefont {Rossini}},
  \bibinfo {author} {\bibfnamefont {R.}~\bibnamefont {Fazio}}, \bibinfo
  {author} {\bibfnamefont {S.}~\bibnamefont {Diehl}}, \ and\ \bibinfo {author}
  {\bibfnamefont {L.}~\bibnamefont {Mazza}},\ }\href@noop {} {\bibfield
  {journal} {\bibinfo  {journal} {Physical Review B}\ }\textbf {\bibinfo
  {volume} {93}},\ \bibinfo {pages} {115113} (\bibinfo {year}
  {2016})}\BibitemShut {NoStop}%
\bibitem [{\citenamefont {Lang}\ and\ \citenamefont
  {B{\"u}chler}(2015)}]{lang2015topological}%
  \BibitemOpen
  \bibfield  {author} {\bibinfo {author} {\bibfnamefont {N.}~\bibnamefont
  {Lang}}\ and\ \bibinfo {author} {\bibfnamefont {H.~P.}\ \bibnamefont
  {B{\"u}chler}},\ }\href@noop {} {\bibfield  {journal} {\bibinfo  {journal}
  {Physical Review B}\ }\textbf {\bibinfo {volume} {92}},\ \bibinfo {pages}
  {041118} (\bibinfo {year} {2015})}\BibitemShut {NoStop}%
\bibitem [{\citenamefont {B{\"u}hler}\ \emph {et~al.}(2014)\citenamefont
  {B{\"u}hler}, \citenamefont {Lang}, \citenamefont {Kraus}, \citenamefont
  {M{\"o}ller}, \citenamefont {Huber},\ and\ \citenamefont
  {B{\"u}chler}}]{buhler2014majorana}%
  \BibitemOpen
  \bibfield  {author} {\bibinfo {author} {\bibfnamefont {A.}~\bibnamefont
  {B{\"u}hler}}, \bibinfo {author} {\bibfnamefont {N.}~\bibnamefont {Lang}},
  \bibinfo {author} {\bibfnamefont {C.~V.}\ \bibnamefont {Kraus}}, \bibinfo
  {author} {\bibfnamefont {G.}~\bibnamefont {M{\"o}ller}}, \bibinfo {author}
  {\bibfnamefont {S.~D.}\ \bibnamefont {Huber}}, \ and\ \bibinfo {author}
  {\bibfnamefont {H.-P.}\ \bibnamefont {B{\"u}chler}},\ }\href@noop {}
  {\bibfield  {journal} {\bibinfo  {journal} {Nature communications}\ }\textbf
  {\bibinfo {volume} {5}},\ \bibinfo {pages} {4504} (\bibinfo {year}
  {2014})}\BibitemShut {NoStop}%
\bibitem [{\citenamefont {Diehl}\ \emph {et~al.}(2008)\citenamefont {Diehl},
  \citenamefont {Micheli}, \citenamefont {Kantian}, \citenamefont {Kraus},
  \citenamefont {B{\"u}chler},\ and\ \citenamefont
  {Zoller}}]{diehl2008quantum}%
  \BibitemOpen
  \bibfield  {author} {\bibinfo {author} {\bibfnamefont {S.}~\bibnamefont
  {Diehl}}, \bibinfo {author} {\bibfnamefont {A.}~\bibnamefont {Micheli}},
  \bibinfo {author} {\bibfnamefont {A.}~\bibnamefont {Kantian}}, \bibinfo
  {author} {\bibfnamefont {B.}~\bibnamefont {Kraus}}, \bibinfo {author}
  {\bibfnamefont {H.}~\bibnamefont {B{\"u}chler}}, \ and\ \bibinfo {author}
  {\bibfnamefont {P.}~\bibnamefont {Zoller}},\ }\href@noop {} {\bibfield
  {journal} {\bibinfo  {journal} {Nature Physics}\ }\textbf {\bibinfo {volume}
  {4}},\ \bibinfo {pages} {878} (\bibinfo {year} {2008})}\BibitemShut {NoStop}%
\bibitem [{\citenamefont {Diehl}\ \emph {et~al.}(2011)\citenamefont {Diehl},
  \citenamefont {Rico}, \citenamefont {Baranov},\ and\ \citenamefont
  {Zoller}}]{diehl2011topology}%
  \BibitemOpen
  \bibfield  {author} {\bibinfo {author} {\bibfnamefont {S.}~\bibnamefont
  {Diehl}}, \bibinfo {author} {\bibfnamefont {E.}~\bibnamefont {Rico}},
  \bibinfo {author} {\bibfnamefont {M.~A.}\ \bibnamefont {Baranov}}, \ and\
  \bibinfo {author} {\bibfnamefont {P.}~\bibnamefont {Zoller}},\ }\href@noop {}
  {\bibfield  {journal} {\bibinfo  {journal} {Nature physics}\ }\textbf
  {\bibinfo {volume} {7}},\ \bibinfo {pages} {971} (\bibinfo {year}
  {2011})}\BibitemShut {NoStop}%
\bibitem [{\citenamefont {Jiang}\ \emph {et~al.}(2011)\citenamefont {Jiang},
  \citenamefont {Kitagawa}, \citenamefont {Alicea}, \citenamefont {Akhmerov},
  \citenamefont {Pekker}, \citenamefont {Refael}, \citenamefont {Cirac},
  \citenamefont {Demler}, \citenamefont {Lukin},\ and\ \citenamefont
  {Zoller}}]{jiang2011majorana}%
  \BibitemOpen
  \bibfield  {author} {\bibinfo {author} {\bibfnamefont {L.}~\bibnamefont
  {Jiang}}, \bibinfo {author} {\bibfnamefont {T.}~\bibnamefont {Kitagawa}},
  \bibinfo {author} {\bibfnamefont {J.}~\bibnamefont {Alicea}}, \bibinfo
  {author} {\bibfnamefont {A.}~\bibnamefont {Akhmerov}}, \bibinfo {author}
  {\bibfnamefont {D.}~\bibnamefont {Pekker}}, \bibinfo {author} {\bibfnamefont
  {G.}~\bibnamefont {Refael}}, \bibinfo {author} {\bibfnamefont {J.~I.}\
  \bibnamefont {Cirac}}, \bibinfo {author} {\bibfnamefont {E.}~\bibnamefont
  {Demler}}, \bibinfo {author} {\bibfnamefont {M.~D.}\ \bibnamefont {Lukin}}, \
  and\ \bibinfo {author} {\bibfnamefont {P.}~\bibnamefont {Zoller}},\
  }\href@noop {} {\bibfield  {journal} {\bibinfo  {journal} {Physical review
  letters}\ }\textbf {\bibinfo {volume} {106}},\ \bibinfo {pages} {220402}
  (\bibinfo {year} {2011})}\BibitemShut {NoStop}%
\bibitem [{\citenamefont {Schuckert}\ \emph {et~al.}(2024)\citenamefont
  {Schuckert}, \citenamefont {Crane}, \citenamefont {Gorshkov}, \citenamefont
  {Hafezi},\ and\ \citenamefont
  {Gullans}}]{schuckert2024fermionqubitfaulttolerantquantumcomputing}%
  \BibitemOpen
  \bibfield  {author} {\bibinfo {author} {\bibfnamefont {A.}~\bibnamefont
  {Schuckert}}, \bibinfo {author} {\bibfnamefont {E.}~\bibnamefont {Crane}},
  \bibinfo {author} {\bibfnamefont {A.~V.}\ \bibnamefont {Gorshkov}}, \bibinfo
  {author} {\bibfnamefont {M.}~\bibnamefont {Hafezi}}, \ and\ \bibinfo {author}
  {\bibfnamefont {M.~J.}\ \bibnamefont {Gullans}},\ }\href
  {https://arxiv.org/abs/2411.08955} {\enquote {\bibinfo {title} {Fermion-qubit
  fault-tolerant quantum computing},}\ } (\bibinfo {year} {2024}),\ \Eprint
  {http://arxiv.org/abs/2411.08955} {arXiv:2411.08955 [quant-ph]} \BibitemShut
  {NoStop}%
\bibitem [{\citenamefont {Holland}\ \emph {et~al.}(2001)\citenamefont
  {Holland}, \citenamefont {Kokkelmans}, \citenamefont {Chiofalo},\ and\
  \citenamefont {Walser}}]{holland2001resonance}%
  \BibitemOpen
  \bibfield  {author} {\bibinfo {author} {\bibfnamefont {M.}~\bibnamefont
  {Holland}}, \bibinfo {author} {\bibfnamefont {S.}~\bibnamefont {Kokkelmans}},
  \bibinfo {author} {\bibfnamefont {M.~L.}\ \bibnamefont {Chiofalo}}, \ and\
  \bibinfo {author} {\bibfnamefont {R.}~\bibnamefont {Walser}},\ }\href@noop {}
  {\bibfield  {journal} {\bibinfo  {journal} {Physical review letters}\
  }\textbf {\bibinfo {volume} {87}},\ \bibinfo {pages} {120406} (\bibinfo
  {year} {2001})}\BibitemShut {NoStop}%
\bibitem [{Note1()}]{Note1}%
  \BibitemOpen
  \bibinfo {note} {In fact, our construction generalizes a subtle discussion of
  optical coherence~\cite {molmer1997optical} to fermionic system}\BibitemShut
  {NoStop}%
\bibitem [{\citenamefont {Gross}\ and\ \citenamefont
  {Bloch}(2017)}]{gross2017quantum}%
  \BibitemOpen
  \bibfield  {author} {\bibinfo {author} {\bibfnamefont {C.}~\bibnamefont
  {Gross}}\ and\ \bibinfo {author} {\bibfnamefont {I.}~\bibnamefont {Bloch}},\
  }\href@noop {} {\bibfield  {journal} {\bibinfo  {journal} {Science}\ }\textbf
  {\bibinfo {volume} {357}},\ \bibinfo {pages} {995} (\bibinfo {year}
  {2017})}\BibitemShut {NoStop}%
\bibitem [{\citenamefont {Bluvstein}\ \emph {et~al.}(2022)\citenamefont
  {Bluvstein}, \citenamefont {Levine}, \citenamefont {Semeghini}, \citenamefont
  {Wang}, \citenamefont {Ebadi}, \citenamefont {Kalinowski}, \citenamefont
  {Keesling}, \citenamefont {Maskara}, \citenamefont {Pichler}, \citenamefont
  {Greiner} \emph {et~al.}}]{bluvstein2022quantum}%
  \BibitemOpen
  \bibfield  {author} {\bibinfo {author} {\bibfnamefont {D.}~\bibnamefont
  {Bluvstein}}, \bibinfo {author} {\bibfnamefont {H.}~\bibnamefont {Levine}},
  \bibinfo {author} {\bibfnamefont {G.}~\bibnamefont {Semeghini}}, \bibinfo
  {author} {\bibfnamefont {T.~T.}\ \bibnamefont {Wang}}, \bibinfo {author}
  {\bibfnamefont {S.}~\bibnamefont {Ebadi}}, \bibinfo {author} {\bibfnamefont
  {M.}~\bibnamefont {Kalinowski}}, \bibinfo {author} {\bibfnamefont
  {A.}~\bibnamefont {Keesling}}, \bibinfo {author} {\bibfnamefont
  {N.}~\bibnamefont {Maskara}}, \bibinfo {author} {\bibfnamefont
  {H.}~\bibnamefont {Pichler}}, \bibinfo {author} {\bibfnamefont
  {M.}~\bibnamefont {Greiner}},  \emph {et~al.},\ }\href@noop {} {\bibfield
  {journal} {\bibinfo  {journal} {Nature}\ }\textbf {\bibinfo {volume} {604}},\
  \bibinfo {pages} {451} (\bibinfo {year} {2022})}\BibitemShut {NoStop}%
\bibitem [{\citenamefont {Kaufman}\ and\ \citenamefont
  {Ni}(2021)}]{Kaufman_2021}%
  \BibitemOpen
  \bibfield  {author} {\bibinfo {author} {\bibfnamefont {A.~M.}\ \bibnamefont
  {Kaufman}}\ and\ \bibinfo {author} {\bibfnamefont {K.-K.}\ \bibnamefont
  {Ni}},\ }\href {\doibase 10.1038/s41567-021-01357-2} {\bibfield  {journal}
  {\bibinfo  {journal} {Nature Physics}\ }\textbf {\bibinfo {volume} {17}},\
  \bibinfo {pages} {1324} (\bibinfo {year} {2021})}\BibitemShut {NoStop}%
\bibitem [{\citenamefont {Kaufman}\ \emph {et~al.}(2014)\citenamefont
  {Kaufman}, \citenamefont {Lester}, \citenamefont {Reynolds}, \citenamefont
  {Wall}, \citenamefont {Foss-Feig}, \citenamefont {Hazzard}, \citenamefont
  {Rey},\ and\ \citenamefont {Regal}}]{Kaufman_2014}%
  \BibitemOpen
  \bibfield  {author} {\bibinfo {author} {\bibfnamefont {A.~M.}\ \bibnamefont
  {Kaufman}}, \bibinfo {author} {\bibfnamefont {B.~J.}\ \bibnamefont {Lester}},
  \bibinfo {author} {\bibfnamefont {C.~M.}\ \bibnamefont {Reynolds}}, \bibinfo
  {author} {\bibfnamefont {M.~L.}\ \bibnamefont {Wall}}, \bibinfo {author}
  {\bibfnamefont {M.}~\bibnamefont {Foss-Feig}}, \bibinfo {author}
  {\bibfnamefont {K.~R.~A.}\ \bibnamefont {Hazzard}}, \bibinfo {author}
  {\bibfnamefont {A.~M.}\ \bibnamefont {Rey}}, \ and\ \bibinfo {author}
  {\bibfnamefont {C.~A.}\ \bibnamefont {Regal}},\ }\href {\doibase
  10.1126/science.1250057} {\bibfield  {journal} {\bibinfo  {journal}
  {Science}\ }\textbf {\bibinfo {volume} {345}},\ \bibinfo {pages} {306}
  (\bibinfo {year} {2014})}\BibitemShut {NoStop}%
\bibitem [{\citenamefont {Levine}\ \emph {et~al.}(2019)\citenamefont {Levine},
  \citenamefont {Keesling}, \citenamefont {Semeghini}, \citenamefont {Omran},
  \citenamefont {Wang}, \citenamefont {Ebadi}, \citenamefont {Bernien},
  \citenamefont {Greiner}, \citenamefont {Vuleti{\'c}}, \citenamefont {Pichler}
  \emph {et~al.}}]{levine2019parallel}%
  \BibitemOpen
  \bibfield  {author} {\bibinfo {author} {\bibfnamefont {H.}~\bibnamefont
  {Levine}}, \bibinfo {author} {\bibfnamefont {A.}~\bibnamefont {Keesling}},
  \bibinfo {author} {\bibfnamefont {G.}~\bibnamefont {Semeghini}}, \bibinfo
  {author} {\bibfnamefont {A.}~\bibnamefont {Omran}}, \bibinfo {author}
  {\bibfnamefont {T.~T.}\ \bibnamefont {Wang}}, \bibinfo {author}
  {\bibfnamefont {S.}~\bibnamefont {Ebadi}}, \bibinfo {author} {\bibfnamefont
  {H.}~\bibnamefont {Bernien}}, \bibinfo {author} {\bibfnamefont
  {M.}~\bibnamefont {Greiner}}, \bibinfo {author} {\bibfnamefont
  {V.}~\bibnamefont {Vuleti{\'c}}}, \bibinfo {author} {\bibfnamefont
  {H.}~\bibnamefont {Pichler}},  \emph {et~al.},\ }\href@noop {} {\bibfield
  {journal} {\bibinfo  {journal} {Physical review letters}\ }\textbf {\bibinfo
  {volume} {123}},\ \bibinfo {pages} {170503} (\bibinfo {year}
  {2019})}\BibitemShut {NoStop}%
\bibitem [{\citenamefont {Evered}\ \emph {et~al.}(2023)\citenamefont {Evered},
  \citenamefont {Bluvstein}, \citenamefont {Kalinowski}, \citenamefont {Ebadi},
  \citenamefont {Manovitz}, \citenamefont {Zhou}, \citenamefont {Li},
  \citenamefont {Geim}, \citenamefont {Wang}, \citenamefont {Maskara} \emph
  {et~al.}}]{evered2023high}%
  \BibitemOpen
  \bibfield  {author} {\bibinfo {author} {\bibfnamefont {S.~J.}\ \bibnamefont
  {Evered}}, \bibinfo {author} {\bibfnamefont {D.}~\bibnamefont {Bluvstein}},
  \bibinfo {author} {\bibfnamefont {M.}~\bibnamefont {Kalinowski}}, \bibinfo
  {author} {\bibfnamefont {S.}~\bibnamefont {Ebadi}}, \bibinfo {author}
  {\bibfnamefont {T.}~\bibnamefont {Manovitz}}, \bibinfo {author}
  {\bibfnamefont {H.}~\bibnamefont {Zhou}}, \bibinfo {author} {\bibfnamefont
  {S.~H.}\ \bibnamefont {Li}}, \bibinfo {author} {\bibfnamefont {A.~A.}\
  \bibnamefont {Geim}}, \bibinfo {author} {\bibfnamefont {T.~T.}\ \bibnamefont
  {Wang}}, \bibinfo {author} {\bibfnamefont {N.}~\bibnamefont {Maskara}},
  \emph {et~al.},\ }\href@noop {} {\bibfield  {journal} {\bibinfo  {journal}
  {Nature}\ }\textbf {\bibinfo {volume} {622}},\ \bibinfo {pages} {268}
  (\bibinfo {year} {2023})}\BibitemShut {NoStop}%
\bibitem [{\citenamefont {Madjarov}\ \emph {et~al.}(2020)\citenamefont
  {Madjarov}, \citenamefont {Covey}, \citenamefont {Shaw}, \citenamefont
  {Choi}, \citenamefont {Kale}, \citenamefont {Cooper}, \citenamefont
  {Pichler}, \citenamefont {Schkolnik}, \citenamefont {Williams},\ and\
  \citenamefont {Endres}}]{madjarov2020high}%
  \BibitemOpen
  \bibfield  {author} {\bibinfo {author} {\bibfnamefont {I.~S.}\ \bibnamefont
  {Madjarov}}, \bibinfo {author} {\bibfnamefont {J.~P.}\ \bibnamefont {Covey}},
  \bibinfo {author} {\bibfnamefont {A.~L.}\ \bibnamefont {Shaw}}, \bibinfo
  {author} {\bibfnamefont {J.}~\bibnamefont {Choi}}, \bibinfo {author}
  {\bibfnamefont {A.}~\bibnamefont {Kale}}, \bibinfo {author} {\bibfnamefont
  {A.}~\bibnamefont {Cooper}}, \bibinfo {author} {\bibfnamefont
  {H.}~\bibnamefont {Pichler}}, \bibinfo {author} {\bibfnamefont
  {V.}~\bibnamefont {Schkolnik}}, \bibinfo {author} {\bibfnamefont {J.~R.}\
  \bibnamefont {Williams}}, \ and\ \bibinfo {author} {\bibfnamefont
  {M.}~\bibnamefont {Endres}},\ }\href@noop {} {\bibfield  {journal} {\bibinfo
  {journal} {Nature Physics}\ }\textbf {\bibinfo {volume} {16}},\ \bibinfo
  {pages} {857} (\bibinfo {year} {2020})}\BibitemShut {NoStop}%
\bibitem [{\citenamefont {Ma}\ \emph {et~al.}(2023)\citenamefont {Ma},
  \citenamefont {Liu}, \citenamefont {Peng}, \citenamefont {Zhang},
  \citenamefont {Jandura}, \citenamefont {Claes}, \citenamefont {Burgers},
  \citenamefont {Pupillo}, \citenamefont {Puri},\ and\ \citenamefont
  {Thompson}}]{ma2023high}%
  \BibitemOpen
  \bibfield  {author} {\bibinfo {author} {\bibfnamefont {S.}~\bibnamefont
  {Ma}}, \bibinfo {author} {\bibfnamefont {G.}~\bibnamefont {Liu}}, \bibinfo
  {author} {\bibfnamefont {P.}~\bibnamefont {Peng}}, \bibinfo {author}
  {\bibfnamefont {B.}~\bibnamefont {Zhang}}, \bibinfo {author} {\bibfnamefont
  {S.}~\bibnamefont {Jandura}}, \bibinfo {author} {\bibfnamefont
  {J.}~\bibnamefont {Claes}}, \bibinfo {author} {\bibfnamefont {A.~P.}\
  \bibnamefont {Burgers}}, \bibinfo {author} {\bibfnamefont {G.}~\bibnamefont
  {Pupillo}}, \bibinfo {author} {\bibfnamefont {S.}~\bibnamefont {Puri}}, \
  and\ \bibinfo {author} {\bibfnamefont {J.~D.}\ \bibnamefont {Thompson}},\
  }\href@noop {} {\bibfield  {journal} {\bibinfo  {journal} {Nature}\ }\textbf
  {\bibinfo {volume} {622}},\ \bibinfo {pages} {279} (\bibinfo {year}
  {2023})}\BibitemShut {NoStop}%
\bibitem [{\citenamefont {Wilk}\ \emph {et~al.}(2010)\citenamefont {Wilk},
  \citenamefont {Ga{\"e}tan}, \citenamefont {Evellin}, \citenamefont {Wolters},
  \citenamefont {Miroshnychenko}, \citenamefont {Grangier},\ and\ \citenamefont
  {Browaeys}}]{wilk2010entanglement}%
  \BibitemOpen
  \bibfield  {author} {\bibinfo {author} {\bibfnamefont {T.}~\bibnamefont
  {Wilk}}, \bibinfo {author} {\bibfnamefont {A.}~\bibnamefont {Ga{\"e}tan}},
  \bibinfo {author} {\bibfnamefont {C.}~\bibnamefont {Evellin}}, \bibinfo
  {author} {\bibfnamefont {J.}~\bibnamefont {Wolters}}, \bibinfo {author}
  {\bibfnamefont {Y.}~\bibnamefont {Miroshnychenko}}, \bibinfo {author}
  {\bibfnamefont {P.}~\bibnamefont {Grangier}}, \ and\ \bibinfo {author}
  {\bibfnamefont {A.}~\bibnamefont {Browaeys}},\ }\href@noop {} {\bibfield
  {journal} {\bibinfo  {journal} {Physical review letters}\ }\textbf {\bibinfo
  {volume} {104}},\ \bibinfo {pages} {010502} (\bibinfo {year}
  {2010})}\BibitemShut {NoStop}%
\bibitem [{\citenamefont {Isenhower}\ \emph {et~al.}(2010)\citenamefont
  {Isenhower}, \citenamefont {Urban}, \citenamefont {Zhang}, \citenamefont
  {Gill}, \citenamefont {Henage}, \citenamefont {Johnson}, \citenamefont
  {Walker},\ and\ \citenamefont {Saffman}}]{isenhower2010demonstration}%
  \BibitemOpen
  \bibfield  {author} {\bibinfo {author} {\bibfnamefont {L.}~\bibnamefont
  {Isenhower}}, \bibinfo {author} {\bibfnamefont {E.}~\bibnamefont {Urban}},
  \bibinfo {author} {\bibfnamefont {X.}~\bibnamefont {Zhang}}, \bibinfo
  {author} {\bibfnamefont {A.}~\bibnamefont {Gill}}, \bibinfo {author}
  {\bibfnamefont {T.}~\bibnamefont {Henage}}, \bibinfo {author} {\bibfnamefont
  {T.~A.}\ \bibnamefont {Johnson}}, \bibinfo {author} {\bibfnamefont
  {T.}~\bibnamefont {Walker}}, \ and\ \bibinfo {author} {\bibfnamefont
  {M.}~\bibnamefont {Saffman}},\ }\href@noop {} {\bibfield  {journal} {\bibinfo
   {journal} {Physical review letters}\ }\textbf {\bibinfo {volume} {104}},\
  \bibinfo {pages} {010503} (\bibinfo {year} {2010})}\BibitemShut {NoStop}%
\bibitem [{\citenamefont {Young}\ \emph {et~al.}(2022)\citenamefont {Young},
  \citenamefont {Eckner}, \citenamefont {Schine}, \citenamefont {Childs},\ and\
  \citenamefont {Kaufman}}]{Young_2022}%
  \BibitemOpen
  \bibfield  {author} {\bibinfo {author} {\bibfnamefont {A.~W.}\ \bibnamefont
  {Young}}, \bibinfo {author} {\bibfnamefont {W.~J.}\ \bibnamefont {Eckner}},
  \bibinfo {author} {\bibfnamefont {N.}~\bibnamefont {Schine}}, \bibinfo
  {author} {\bibfnamefont {A.~M.}\ \bibnamefont {Childs}}, \ and\ \bibinfo
  {author} {\bibfnamefont {A.~M.}\ \bibnamefont {Kaufman}},\ }\href {\doibase
  10.1126/science.abo0608} {\bibfield  {journal} {\bibinfo  {journal}
  {Science}\ }\textbf {\bibinfo {volume} {377}},\ \bibinfo {pages} {885}
  (\bibinfo {year} {2022})}\BibitemShut {NoStop}%
\bibitem [{\citenamefont {Tao}\ \emph {et~al.}(2024)\citenamefont {Tao},
  \citenamefont {Ammenwerth}, \citenamefont {Gyger}, \citenamefont {Bloch},\
  and\ \citenamefont {Zeiher}}]{Tao_2024}%
  \BibitemOpen
  \bibfield  {author} {\bibinfo {author} {\bibfnamefont {R.}~\bibnamefont
  {Tao}}, \bibinfo {author} {\bibfnamefont {M.}~\bibnamefont {Ammenwerth}},
  \bibinfo {author} {\bibfnamefont {F.}~\bibnamefont {Gyger}}, \bibinfo
  {author} {\bibfnamefont {I.}~\bibnamefont {Bloch}}, \ and\ \bibinfo {author}
  {\bibfnamefont {J.}~\bibnamefont {Zeiher}},\ }\href {\doibase
  10.1103/PhysRevLett.133.013401} {\bibfield  {journal} {\bibinfo  {journal}
  {Phys. Rev. Lett.}\ }\textbf {\bibinfo {volume} {133}},\ \bibinfo {pages}
  {013401} (\bibinfo {year} {2024})}\BibitemShut {NoStop}%
\bibitem [{Note2()}]{Note2}%
  \BibitemOpen
  \bibinfo {note} {For simplicity we assume $M_r> N>M_s$, but discuss
  generalizations in SM}\BibitemShut {NoStop}%
\bibitem [{Note3()}]{Note3}%
  \BibitemOpen
  \bibinfo {note} {We note that the entire construction here is employed for
  fermions, but can be applied to bosons as well.}\BibitemShut {Stop}%
\bibitem [{Note4()}]{Note4}%
  \BibitemOpen
  \bibinfo {note} {We define the boundary term in \protect \textup {\hbox
  {\mathsurround \z@ \protect \normalfont (\ignorespaces \ref {eq:R}\unskip
  \@@italiccorr )}} via $\eta _{0}=1$, $\eta _{M_r+1}=0$.}\BibitemShut {Stop}%
\bibitem [{Note5()}]{Note5}%
  \BibitemOpen
  \bibinfo {note} {This particular set of reference states is related to the
  choice of $R$. It is perfectly possible to generalize this and choose
  different reference states and correspondingly $R$.}\BibitemShut {Stop}%
\bibitem [{Note6()}]{Note6}%
  \BibitemOpen
  \bibinfo {note} {Our construction can interpreted as encoding multiple
  logical modes in a single code block, evading the no-go theorem of Ref.~\cite
  {schuckert2024fermionqubitfaulttolerantquantumcomputing}.}\BibitemShut
  {Stop}%
\bibitem [{\citenamefont {Knill}\ and\ \citenamefont
  {Laflamme}(1997)}]{knill1997theory}%
  \BibitemOpen
  \bibfield  {author} {\bibinfo {author} {\bibfnamefont {E.}~\bibnamefont
  {Knill}}\ and\ \bibinfo {author} {\bibfnamefont {R.}~\bibnamefont
  {Laflamme}},\ }\href@noop {} {\bibfield  {journal} {\bibinfo  {journal}
  {Physical Review A}\ }\textbf {\bibinfo {volume} {55}},\ \bibinfo {pages}
  {900} (\bibinfo {year} {1997})}\BibitemShut {NoStop}%
\bibitem [{Note7()}]{Note7}%
  \BibitemOpen
  \bibinfo {note} {We assume that these ancilla qubits are error-free as they
  can in principle be encoded and error corrected using standard
  techniques~\cite {gottesman1997stabilizer}.}\BibitemShut {Stop}%
\bibitem [{Note8()}]{Note8}%
  \BibitemOpen
  \bibinfo {note} {Error correction for multiple blocks can be essentially
  parallelized.}\BibitemShut {Stop}%
\bibitem [{Note9()}]{Note9}%
  \BibitemOpen
  \bibinfo {note} {In the SM we illustrate this on the example of a fermionic
  Steane code.}\BibitemShut {Stop}%
\bibitem [{\citenamefont {Bayha}\ \emph {et~al.}(2020)\citenamefont {Bayha},
  \citenamefont {Holten}, \citenamefont {Klemt}, \citenamefont {Subramanian},
  \citenamefont {Bjerlin}, \citenamefont {Reimann}, \citenamefont {Bruun},
  \citenamefont {Preiss},\ and\ \citenamefont {Jochim}}]{bayha2020observing}%
  \BibitemOpen
  \bibfield  {author} {\bibinfo {author} {\bibfnamefont {L.}~\bibnamefont
  {Bayha}}, \bibinfo {author} {\bibfnamefont {M.}~\bibnamefont {Holten}},
  \bibinfo {author} {\bibfnamefont {R.}~\bibnamefont {Klemt}}, \bibinfo
  {author} {\bibfnamefont {K.}~\bibnamefont {Subramanian}}, \bibinfo {author}
  {\bibfnamefont {J.}~\bibnamefont {Bjerlin}}, \bibinfo {author} {\bibfnamefont
  {S.~M.}\ \bibnamefont {Reimann}}, \bibinfo {author} {\bibfnamefont {G.~M.}\
  \bibnamefont {Bruun}}, \bibinfo {author} {\bibfnamefont {P.~M.}\ \bibnamefont
  {Preiss}}, \ and\ \bibinfo {author} {\bibfnamefont {S.}~\bibnamefont
  {Jochim}},\ }\href@noop {} {\bibfield  {journal} {\bibinfo  {journal}
  {Nature}\ }\textbf {\bibinfo {volume} {587}},\ \bibinfo {pages} {583}
  (\bibinfo {year} {2020})}\BibitemShut {NoStop}%
\bibitem [{\citenamefont {Hensgens}\ \emph {et~al.}(2017)\citenamefont
  {Hensgens}, \citenamefont {Fujita}, \citenamefont {Janssen}, \citenamefont
  {Li}, \citenamefont {Van~Diepen}, \citenamefont {Reichl}, \citenamefont
  {Wegscheider}, \citenamefont {Das~Sarma},\ and\ \citenamefont
  {Vandersypen}}]{hensgens2017quantum}%
  \BibitemOpen
  \bibfield  {author} {\bibinfo {author} {\bibfnamefont {T.}~\bibnamefont
  {Hensgens}}, \bibinfo {author} {\bibfnamefont {T.}~\bibnamefont {Fujita}},
  \bibinfo {author} {\bibfnamefont {L.}~\bibnamefont {Janssen}}, \bibinfo
  {author} {\bibfnamefont {X.}~\bibnamefont {Li}}, \bibinfo {author}
  {\bibfnamefont {C.}~\bibnamefont {Van~Diepen}}, \bibinfo {author}
  {\bibfnamefont {C.}~\bibnamefont {Reichl}}, \bibinfo {author} {\bibfnamefont
  {W.}~\bibnamefont {Wegscheider}}, \bibinfo {author} {\bibfnamefont
  {S.}~\bibnamefont {Das~Sarma}}, \ and\ \bibinfo {author} {\bibfnamefont
  {L.~M.}\ \bibnamefont {Vandersypen}},\ }\href@noop {} {\bibfield  {journal}
  {\bibinfo  {journal} {Nature}\ }\textbf {\bibinfo {volume} {548}},\ \bibinfo
  {pages} {70} (\bibinfo {year} {2017})}\BibitemShut {NoStop}%
\bibitem [{\citenamefont {Maskara}\ \emph {et~al.}(2023)\citenamefont
  {Maskara}, \citenamefont {Ostermann}, \citenamefont {Shee}, \citenamefont
  {Kalinowski}, \citenamefont {Gomez}, \citenamefont {Bravo}, \citenamefont
  {Wang}, \citenamefont {Krylov}, \citenamefont {Yao}, \citenamefont
  {Head-Gordon} \emph {et~al.}}]{maskara2023programmable}%
  \BibitemOpen
  \bibfield  {author} {\bibinfo {author} {\bibfnamefont {N.}~\bibnamefont
  {Maskara}}, \bibinfo {author} {\bibfnamefont {S.}~\bibnamefont {Ostermann}},
  \bibinfo {author} {\bibfnamefont {J.}~\bibnamefont {Shee}}, \bibinfo {author}
  {\bibfnamefont {M.}~\bibnamefont {Kalinowski}}, \bibinfo {author}
  {\bibfnamefont {A.~M.}\ \bibnamefont {Gomez}}, \bibinfo {author}
  {\bibfnamefont {R.~A.}\ \bibnamefont {Bravo}}, \bibinfo {author}
  {\bibfnamefont {D.~S.}\ \bibnamefont {Wang}}, \bibinfo {author}
  {\bibfnamefont {A.~I.}\ \bibnamefont {Krylov}}, \bibinfo {author}
  {\bibfnamefont {N.~Y.}\ \bibnamefont {Yao}}, \bibinfo {author} {\bibfnamefont
  {M.}~\bibnamefont {Head-Gordon}},  \emph {et~al.},\ }\href@noop {} {\bibfield
   {journal} {\bibinfo  {journal} {arXiv preprint arXiv:2312.02265}\ }
  (\bibinfo {year} {2023})}\BibitemShut {NoStop}%
\bibitem [{\citenamefont {M{\o}lmer}(1997)}]{molmer1997optical}%
  \BibitemOpen
  \bibfield  {author} {\bibinfo {author} {\bibfnamefont {K.}~\bibnamefont
  {M{\o}lmer}},\ }\href@noop {} {\bibfield  {journal} {\bibinfo  {journal}
  {Physical Review A}\ }\textbf {\bibinfo {volume} {55}},\ \bibinfo {pages}
  {3195} (\bibinfo {year} {1997})}\BibitemShut {NoStop}%
\end{thebibliography}%

\appendix
\clearpage

\onecolumngrid
\section*{Supplemental Material to\\ ``Error-corrected fermionic quantum processors with neutral atoms''}

In this supplemental material, we provide detailed calculations and explanations for the reference and code constructions described in the main text.\\

\twocolumngrid

\subsection{Details on the fermionic reference}

\subsubsection{Hilbert space}
We start by recalling the definition of the Hilbert space $\mathcal{H}$. It is spanned by the states $\ket{\Omega}\equiv r_N^\dag r_{N-1}^\dag ..r^\dag_1\ket{\rm vac}$ and all states that can be generated from it by application of the creation operator of referenced fermions~$c_i^\dag=s_i^\dag R$. 
We note that for this definition we require $N\leq M_r$. Moreover, if the number of system modes $M_s\leq N$, then $\mathcal{H}$ has the structure of a full fermionic Fock space over $M_s$ modes.

Note, that for a given $M_s$ the minimal number of physical particles $N$ and reference modes $M_r$ required to obtain the full fermionic Fock space is $N=M_r=M_s$. By generalizing the above construction one can also get subspaces of the fermionic Fock spaces with both lower and upper caps on the total number of referenced fermions. Specifically, if $N>M_r$ one can have a Fock space with a minimum number $N-M_r$ of referenced fermions, while for $N<M_s$ the maximum number is $N$. 

As discussed in the main text, all states on $\mathcal{H}$ have a simple structure on the reference. Specifically, the reference can be only in one of $N+1$ different reference states
\begin{align}
    \ket{N-n}_R \equiv R^n\ket{\Omega} =  r_{N-n}^\dag \dots r^\dag_1\ket{\rm vac}\, ,
\end{align}
with $n=0,1,\dots N$.
Specifically, they read
\begin{align}
\label{eq:reference-states}
    \ket{N}_R & = \phantom{R^2}\ket{\Omega}=   r_{N}^\dag r_{N-1}^\dag r_{N-2}^\dag \dots r^\dag_1\ket{\rm vac} \nonumber\\
    \ket{N-1}_R & = R\phantom{^2}\ket{\Omega}=   \phantom{r_{N}^\dag}r_{N-1}^\dag r_{N-2}^\dag \dots r^\dag_1\ket{\rm vac}\nonumber\\
    \ket{N-2}_R & = R^2\ket{\Omega} = \phantom{r_{N}^\dag r_{N-1}^\dag}  r_{N-2}^\dag \dots r^\dag_1\ket{\rm vac}\nonumber\\
    &\vdots\nonumber\\
    \ket{0}_R &= R^N\ket{\Omega}= \phantom{r_{N}^\dag r_{N-1}^\dag r_{N-2}^\dag \dots r^\dag_1}\ket{\rm vac}\, .
\end{align}
Importantly, for all states in $\mathcal{H}$ with $n$ referenced fermions, the reference is in the state $\ket{n}_R$. That is, all eigenstates of $\sum_i c_i^\dag c_i$ with eigenvalue $n$ take the form of $\ket{N-n}_R$ on the reference modes. Note, for $N>M_s$ only the first $M_s+1$ reference states, i.e., the states $\ket{n}_R$ with $n=N,N-1,\dots, N-M_s$, are relevant.

We note that this construction of the reference and the associated Hilbert space of referenced fermions is not unique. For a more general construction note that any set of $M_s+1$ reference states $\ket{n}_R$  that contain $n=N,N-1,\dots, N-M_s$ physical fermions, i.e. $\sum_i r_i^\dag r_i\ket{n}_R=n\ket{n}_R$, could be used to serve as a reference. The ladder operator $R$ simply needs to be defined to act as $R\ket{n}_R=\ket{n-1}_R$. This could be relevant for other physical realizations of a fermionic reference. For the discussion below we use the definition of $R$ given in the main text, but analogous derivations can be carried out with such generalized references.

\subsubsection{Commutation relations and properties of the reference operators}
In this section, we derive the commutation relations for referenced fermion operators. For this we start out by discussing properties of the reference operator $R$.
This will also be useful in the next sections, where we detail the decomposition of the system-reference operations into elementary two-mode unitary gates.
For simplicity in the derivations below we assume $M_r>N>M_s$.

We first start by writing the reference operator $R$ as a sum of local operators $R_j$ as
\begin{align}
\label{eq:local-decomp}
    R = \sum_{j=1}^{M_r} R_j\,,
\end{align}
with
\begin{subequations}
\label{eq:Rj}
    \begin{align}
    R_j &= (1-\eta_{j+1})r_j \eta_{j-1}\,,\\
    R_j^\dagger &= (1-\eta_{j+1})r^\dagger_j \eta_{j-1}\, .
\end{align}
\end{subequations}
here, the boundary terms are defined with $\eta_0 = 1$ and $\eta_{M_r+1} = 0$. The operators $R_j$ trivially satisfy
\begin{subequations}
    \begin{align}
    \label{eq:relations-a}
    &R_j^\dagger R_j = (1-\eta_{j+1})\eta_j \eta_{j-1}\, ,\\
    \label{eq:relations-b}
    &R_j R_j^\dagger =(1-\eta_{j+1})(1-\eta_j) \eta_{j-1} \, ,\\
    & R_{j} R_{j+1}= (1-\eta_{j+2})r_j r_{j+1} \eta_{j-1} \, .
\end{align}
These last identities follow directly from the structure of the reference states.
\end{subequations}
In addition, on the Hilbert space~$\mathcal{H}$ we also have the relations
\begin{subequations}
    \begin{align}
    \label{eq:property-c}
    R_{j\neq i}^\dagger R_{i}\mathbf{P} =\mathbf{P}R_{j\neq i}^\dagger R_{i} = 0\, .
\end{align}
\end{subequations}
They can be easily checked by noting that in $\mathcal{H}$ the reference has to be in one of the $N+1$ reference states given in eq.~\eqref{eq:reference-states}.
Using these identities, we can determine the commutation relations between $R$ and $R^\dagger$ on $\mathcal{H}$, as well as between $R_i$ and $R^\dagger_j$, which we use further below to find a decomposition of the system-reservoir operations on $\mathcal{H}$.

Using the relations \eqref{eq:relations-a}, \eqref{eq:relations-b}, and \eqref{eq:property-c}, we obtain the following expression for the composition of reference operators
\begin{subequations}
\label{eq:composition}
\begin{align}
    R^\dagger R \mathbf{P} &= \sum_{i} R_i^\dagger R_i \mathbf{P}=\sum_i  (1-\eta_{i+1})\eta_i \eta_{i-1}\mathbf{P} \nonumber\\
    &=\sum_i  (1-\eta_{i+1})\eta_i \mathbf{P} \, ,\\
    R R^\dagger\mathbf{P} &= \sum_{i}  R_i R_i^\dagger \mathbf{P}=\sum_i  (1-\eta_{i+1})(1-\eta_i) \eta_{i-1}\mathbf{P}\nonumber\\
    &=\sum_i  (1-\eta_i) \eta_{i-1}\mathbf{P}\, ,
\end{align}
\end{subequations}
as well as analogous equations for $\mathbf{P}R^\dagger R $ and $\mathbf{P}R R^\dagger$. The last equality holds because the structure of $\mathcal{H}$ implies that if a reference mode $i$ is occupied, i.e. $\eta_i=1$, then also every other reference mode $j$ with $j<i$ is occupied as well. Similarly, if the referenced mode $i$ is unoccupied, $\eta_i=0$, then every other reference mode $j>i$ is unoccupied too, i.e. $\eta_{j>i}=0$. Furthermore, we obtain that
\begin{subequations}
\label{eq:identity-on-P}
\begin{align}
    \sum_i  (1-\eta_{i+1})\eta_i \mathbf{P}&= \mathbf{P}\, ,\\
    \sum_i  (1-\eta_{i})\eta_{i-1} \mathbf{P}&= \mathbf{P}\, ,
\end{align}
\end{subequations}
which can be verified explicitly as for any reference state $R^n\ket{\Omega}$ exactly one of the projectors $(1-\eta_{i+1})\eta_i$ is equal to one, while all others are equal to zero. Therefore, using Eq.~\eqref{eq:identity-on-P}, one finds from Eq.~\eqref{eq:composition}
\begin{align}
\label{eq:RdagR}
   R^\dagger R \mathbf{P}&=\mathbf{P}R^\dagger R = \mathbf{P}\\
    R R^\dagger\mathbf{P}&= \mathbf{P} R R^\dagger= \mathbf{P}
   \, .
\end{align}
As a consequence, we obtain the commutation relation
\begin{align}
   [R^\dagger,R]\mathbf{P}= \mathbf{P}[R^\dagger,R] = 0\, .
\end{align}
In contrast the commutation relations of $R$ with the system operators are \emph{fermionic}
\begin{align}
\label{eq:anticomm-R-s}
    \{R,s_i\} = 0\, .
\end{align}
This follows from Eqs.~\eqref{eq:local-decomp}, \eqref{eq:Rj} and the anti-commutation relation $\{r_j,s_i\} = 0$ for all $i,j$. The anti-commutation of the referenced fermions follows as
\begin{align}
\{c^\dagger_i, c_j\} \mathbf{P}&=  s^\dagger_i  R R^\dagger  s_j\mathbf{P} -  R^\dagger s_j s^\dagger_i  R\mathbf{P} \nonumber \\
&= s^\dagger_i s_j  R R^\dagger  \mathbf{P} -  s_j s^\dagger_i R^\dagger  R\mathbf{P}  \nonumber \\
&=  s^\dagger_i s_j    \mathbf{P} -  s_j s^\dagger_i \mathbf{P} \nonumber \\
&=  \{ s^\dagger_i,s_j\}   \mathbf{P}\nonumber\\
&= \delta_{ij}\mathbf{P}  \, .
\end{align}
Here we used the relations \eqref{eq:anticomm-R-s} and~\eqref{eq:RdagR}. Similarly, we also obtain $\mathbf{P}\{c^\dagger_i, c_j\} =\delta_{ij}\mathbf{P}$.

\subsubsection{Decomposition of operations that coherently create referenced fermions}
In this subsection we discuss the decomposition of the operator $e^{i\theta (c_i^\dag+c_i)}$ in terms of physical operations. To prepare for this we first note that for any mode $i$
\begin{align}
    &[R_ks_i^\dagger + \text{\text{H.c.}},R_ls_i^\dagger + \text{\text{H.c.}}]\mathbf{P}\nonumber\\ &=( R_ks_i^\dagger s_i R_l^\dagger + s_i R_k^\dagger R_l s_i^\dagger-(R_ls_i^\dagger s_i R_k^\dagger + s_i R_l^\dagger R_k s_i^\dagger))\mathbf{P}\nonumber\\ &=(s_i^\dagger s_i R_k R_l^\dagger + s_i s_i^\dagger R_k^\dagger R_l -(s_i^\dagger s_i R_lR_k^\dagger + s_i  s_i^\dagger R_l^\dagger R_k))\mathbf{P}\nonumber\\
    &= 0\, ,
\end{align}
which follows from the fermionic commutation relations of the physical modes, as well as the properties of the operators $R_i$ discussed in the previous section. For $k=l$ this expression vanishes trivially as each term gets cancelled by its counterpart. For $k\neq l$, each term vanishes individually due to property~\eqref{eq:property-c}. Hence, we can decompose
\begin{align}
    e^{i\theta (c^\dagger_i + c_i)} \mathbf{P}= e^{i\theta (R^\dagger  s_i  +  s_i^\dagger R ) } \mathbf{P}= \prod_{k=1}^{M_r} e^{i\theta ( R_k^\dagger s_i +  s_i^\dagger R_k) }\mathbf{P} \, .
\end{align}

In the following, we aim to further decompose $e^{i\theta (c^\dagger_i + c_i)}$ into physical two-mode gates. To this end, we first note the following relations for tunneling operations between system modes $s_i$ and reference modes $r_j$ (see also~\cite{gonzalez2023fermionic}). First, consider 
\begin{align}
    e^{i\pi \eta_j \eta_k} (r_j^\dagger s_i + \text{\text{H.c.}})e^{i\pi \eta_j \eta_k} = r_j^\dagger (1-2\eta_k) s_i + \text{\text{H.c.}}
\end{align}
Therefore, a composition of tunneling and density interactions yields
\begin{align}
    &e^{i\theta(r_j^\dagger s_i + \text{\text{H.c.}})} e^{i\pi \eta_j \eta_k} e^{i\theta(r_j^\dagger s_i + \text{\text{H.c.}})}e^{i\pi \eta_j \eta_k}\nonumber\\
    &= e^{i(2\theta)(1-\eta_k)(r_j^\dagger s_i + \text{\text{H.c.}})}\, ,
\end{align}
and similarly
\begin{align}
    &e^{i\theta(r_j^\dagger s_i + \text{\text{H.c.}})} e^{i\pi \eta_j \eta_k} e^{-i\theta(r_j^\dagger s_i + \text{\text{H.c.}})}e^{i\pi \eta_j \eta_k}\nonumber\\
    &= e^{i(2\theta)\eta_k(r_j^\dagger s_i + \text{\text{H.c.}})}\, .
\end{align}
To obtain the desired interaction, consider next (with $k\neq j$ and $l\neq j$)
\begin{align}
    &e^{i\pi \eta_j (\eta_k+\eta_l)} e^{i\theta(r_j^\dagger s_i + \text{\text{H.c.}})}e^{i\pi \eta_j (\eta_k+\eta_l)}\nonumber\\
    &= e^{i\theta(1-2\eta_k)(1-2\eta_l)(r_j^\dagger s_i + \text{\text{H.c.}})}\, .
\end{align}
That is, the tunneling operation obtains an additional phase if $\eta_k$ and $\eta_l$ are different, and it does not obtain a phase if  $\eta_k$ and $\eta_l$ are equal. In the space $\mathcal{H}$, we use this to decompose the system-reference tunneling as
\begin{align}
     e^{i\theta (  R^\dagger s_i +  s_i^\dagger R ) } &= \prod_k  e^{i\frac{\theta}{2}(s_i^\dagger r_k + \text{H.c.})}e^{i\pi \eta_k (\eta_{k+1}+\eta_{k-1})}\nonumber\\
     &\quad \times e^{-i\frac{\theta}{2}(s_i^\dagger r_k + \text{H.c.})}e^{i\pi \eta_k (\eta_{k+1}+\eta_{k-1})}\, ,
\end{align}
which is the expression given in the main text.

We note that, in principle, the system-reference tunneling can be decomposed in different ways, e.g., such that the required physical interactions involving the lattice are limited to physical tunneling of the atoms into and out of tweezers.

\subsection{System error correction}
\subsubsection{Repetition code}
Here, we show that the repetition code given in the main text forms an error-correcting code, by checking the Knill-Laflamme error correcting condition explicitly for the case with a single logical fermionic mode. It is given by
\begin{align}
    \bra{i}_L E^\dagger_n E_m \ket{j}_L = \delta_{ij} \mathcal{C}_{nm}\, 
\end{align}
where $\mathcal{C}$ is a hermitian matrix independent of the code words $i_L$, $j_L$. Recall that,
\begin{subequations}
    \begin{align}
        \ket{0}_L &= \frac{1}{2}( 1 + ic_1^\dagger c_{2}^\dagger-ic_2^\dagger c_{3}^\dagger+ c_1^\dagger c_{3}^\dagger )\ket{\Omega}\, ,\\
        \ket{1}_L &= \frac{1}{2}( ic_1^\dagger c_2^\dagger c_3^\dagger - c_1^\dagger +ic_2^\dagger -c_{3}^\dagger  )\ket{\Omega}\, .
    \end{align}
\end{subequations}
The local particle number in each state is
\begin{align}
    \bra{0}_L n_i \ket{0}_L = \bra{1}_L n_i \ket{1}_L = \frac{1}{2}\, , 
\end{align}
where we used that $s^\dagger_i s_i = c^\dagger_i c_i$. Also, the observable is diagonal in the logical states, i.e.
\begin{align}
    \bra{1}_Ln_i\ket{0}_L = 0\, .
\end{align}
The densities are linearly related with the errors, i.e. $E_i=1-2s^\dagger_i s_i$, and it is easily checked that the error correction condition is fulfilled, such that the dephasing error channel with Kraus operators $K_0 = \sqrt{1-3p}\,\mathbb{1}$ and $K_i = \sqrt{p}(1-2n_i) $ is correctable. A second version of the Knill-Laflamme condition, which we use for the explicit calculations for the Steane code below, is given by~\cite{nielsen2010quantum}
\begin{align}
    P E^\dagger_\alpha E_\beta P = \mathcal{C}_{\alpha\beta} P\, ,
\end{align}
where the projector $P$ onto the code space $\mathcal{H}^C$ is given by
\begin{align}
    P &= \frac{1+S_{12}}{2}\frac{1+S_{23}}{2}\, .
\end{align}
The condition is straightforwardly fulfilled since phase errors anti-commute with at least one of the stabilizers.

The corresponding multi-mode version of the conditions can be checked analogously.

\subsubsection{Steane code to correct for particle loss}
In this section, we extend the discussion of the main text to more general error correcting codes. In particular we give a code that can correct for atom loss. Here we discuss this on the example of a single logical mode encoded in a fermionic version of the Steane code.

\paragraph{Code.--} We employ a construction based on the well-known Steane code. That is, we consider seven referenced fermionic modes with corresponding annihilation operators $c_1,...,c_7$ and the following stabilizers
\begin{subequations}
\begin{align}
    S^Z_{4567}=&(1-2c^\dagger_4 c_4)(1-2c^\dagger_5 c_5)(1-2c^\dagger_6 c_6)(1-2c^\dagger_7 c_7)\nonumber\\
    S^Z_{2367}=&(1-2c^\dagger_2 c_2)(1-2c^\dagger_3 c_3)(1-2c^\dagger_6 c_6)(1-2c^\dagger_7 c_7)\nonumber \\
    S^Z_{1357}=&(1-2c^\dagger_1 c_1)(1-2c^\dagger_3 c_3)(1-2c^\dagger_5 c_5)(1-2c^\dagger_7 c_7)\nonumber\\
S^X_{4567}=&i(c_4+c_4^\dagger)i(c_5+c_5^\dagger)i(c_6+c_6^\dagger)i(c_7+c_7^\dagger)\nonumber\\
S^X_{2367}=&i(c_2+c_2^\dagger)i(c_3+c_3^\dagger)i(c_6+c_6^\dagger)i(c_7+c_7^\dagger)\nonumber\\    S^X_{1357}=&i(c_1+c_1^\dagger)i(c_3+c_3^\dagger)i(c_5+c_5^\dagger)i(c_7+c_7^\dagger)
    \, .
\end{align}    
The projector $P$ onto the code space $\mathcal{H}^C$ is given by
\begin{align}
    P &= \frac{1+S^X_{4567}}{2}\frac{1+S^X_{2367}}{2}\frac{1+S^X_{1357}}{2}\nonumber\\
   &\quad \times\frac{1+S^Z_{4567}}{2}\frac{1+S^Z_{2367}}{2}\frac{1+S^Z_{1357}}{2}\, .
\end{align}

\end{subequations}

\paragraph{Loss channel}
Correction of phase errors $p_i$ for the modes $i=1,..,7$ works analagous to the example of a repetition code discussed above. Here, we consider instead an error channel for fermion loss in the system modes. That is, we define a loss channel with Kraus operators $K_0=\sqrt{1-7p}\,\mathbb{1}$, $K_{i}=\sqrt{p}\, s_i$, $K_{i+7} = \sqrt{p}(1-n_i)$, fulfilling $\sum_\alpha K_\alpha^\dagger K_\alpha = \mathbb{1}$, as
\begin{align}
    \rho \mapsto \sum_{\alpha=0}^{14} K_\alpha \rho K_\alpha^\dagger\, .
\end{align}

\paragraph{Knill-Laflamme condition}
We now demonstrate the fulfillment of the Knill-Laflamme conditions under the above error channel. This means, we demonstrate the conditions $P K_\alpha ^\dagger K_\beta P \propto P$.

Consider first the example $\alpha=\beta=i$. Without loss of generality, we show the error channel for the mode $i=1$. First, consider
\begin{align}
    &\frac{1+S^X_{1357}}{2} K^\dagger_1 K_1 \frac{1+S^X_{1357}}{2} \nonumber\\
    &=  p\frac{1+S^X_{1357}}{2} \frac{1-p_1}{2} \frac{1+S^X_{1357}}{2}\nonumber\\
    &= p\frac{1+S^X_{1357}}{4} - \frac{p_1}{2} p\frac{1-S^X_{1357}}{2}\frac{1+S^X_{1357}}{2} \nonumber\\
    &= p\frac{1+S^X_{1357}}{4}\, ,
\end{align}
where $p_i=1-2n_i$ is the local physical parity operator which anti-commutes with the stabilizer. More generally, from this follows,
\begin{align}
    P K^\dagger_i K_i P  = \frac{p}{2} P\, ,
\end{align}
which is in line with the error correction condition. Consider next the case with $\alpha=i$ and $\beta=j\neq i$. For any combination there is a stabilizer $S^Z$ involving mode $i$ but not mode $j$. For this stabilizer, we get
\begin{align}
    \frac{1+S^Z}{2} K^\dagger_{i} K_j \frac{1+S^Z}{2}&= \frac{1+S^Z}{2}  \frac{1-S^Z}{2} K^\dagger_{i} K_j,\\
    \rightarrow P K^\dagger_{i} K_j P &=0\propto P\, ,
\end{align}
in agreement with the condition. 

Similarly, the condition is fulfilled for the case  $\alpha=0$, and $\beta = j$, and without loss of generality we choose $j=1$. We get
\begin{align}
    &\frac{1+S^Z_{1357}}{2} K^\dagger_0 K_1 \frac{1+S^Z_{1357}}{2} \nonumber\\
    &=  \sqrt{1-p}\sqrt{p} \frac{1+S^Z_{1357}}{2} s_1 \frac{1+S^Z_{1357}}{2}\nonumber\\
    &=  \sqrt{1-p}\sqrt{p} \frac{1+S^Z_{1357}}{2} \frac{1-S^Z_{1357}}{2} s_1 \nonumber\\ &= 0\, ,
\end{align}
and hence, more generally,
\begin{align}
    P K^\dagger_0 K_i P  = 0 \propto P \, .
\end{align}

The correctability of the Kraus operations $K_{i+7}$ follows from the correctability of phase errors, since $K_{i+7} = \sqrt{p}\, [\tfrac{1}{2}p_i +\tfrac{1}{2}]$. For instance, consider $\alpha=i$ and $\beta = j+7$, such that
\begin{align}
    K^\dagger_i K_{j+7} = \frac{\sqrt{p}}{2} K^\dagger_i + \frac{\sqrt{p}}{2} K^\dagger_i p_j\, .
\end{align}
From our previous considerations follows directly
\begin{align}
    P K^\dagger_i K_{j+7} P &= \frac{\sqrt{p}}{2} P K^\dagger_i p_j P\nonumber\\
    &= \frac{\sqrt{p}}{2} P s^\dagger_i p_j P\, .
\end{align}
There is at least one stabilizer $S^Z$ which is flipped by $s_i^\dagger$ but which commutes with $p_j$, i.e.
\begin{align}
    \frac{1+S^Z}{2} s^\dagger_i p_j \frac{1+S^Z}{2} = \frac{1+S^Z}{2}\frac{1-S^Z}{2} s^\dagger_i p_j =0 .
\end{align}
Therefore, the error-correction condition is fulfilled
\begin{align}
    P K^\dagger_i K_{j+7} P = 0\propto P\, .
\end{align}

The error correction conditions can be shown to be fulfilled for all combinations of single-mode errors acting on any of the physical modes of a block. For the losses, stabilizer measurements uniquely identify the lost fermion and subsequently it can be replaced from the reference. Again, the corresponding multi-mode condition follows analogously.

\subsection{Subspace with fixed number of logical fermions:  reference error correction and number conserving logical gates}
In the main text, we discuss the advantages of restricting the logical Hilbert space to the subspace that contains $N_L$ logical fermions in the logical modes, i.e., the subspace denoted by $\mathcal{H}^C_{N_L}$. Here, we detail the following two advantages explicitly. First, logical operations in $\mathcal{H}_{N_L}^C$ can be obtained from operations on the system modes only. Second, measurements on the reference states do not reveal the logical states, which implies that error on the reference modes can be corrected. In addition we discuss physical implementation of logical gates, both transversal and not transversal. 

We first illustrate this on a simple example of two logical modes, and then discuss the general case.

\subsubsection{Simple two-mode example}

We consider our simple fermionic repetition code on two logical modes with a single logical fermion. The logical states are
\begin{subequations}
\begin{align}   
\ket{1,0}_L&= \ket{111}\ket{000} \ket{2}_R\nonumber\\
&+ \big(\ket{001}+\ket{010}+\ket{100}\big)\nonumber\\
&\qquad\big(\ket{110}+\ket{101}+\ket{011}\big)\ket{2}_R\nonumber\\
&+ \ket{111}\big(\ket{110}+\ket{101}+\ket{011}\big)\ket{0}_R\nonumber\\
&+ \big(\ket{001}+\ket{010}+\ket{100}\big)\ket{000} \ket{4}_R\, ,  \\
\ket{0,1}_L&= \ket{000}\ket{111} \ket{2}_R\nonumber\\
&+ \big(\ket{110}+\ket{101}+\ket{011}\big)\nonumber\\
&\qquad\big(\ket{001}+\ket{010}+\ket{100}\big) \ket{2}_R\nonumber\\
&+ \big(\ket{110}+\ket{101}+\ket{011}\big)\ket{111}\ket{0}_R\nonumber\\
&+ \ket{000}\big(\ket{001}+\ket{010}+\ket{100}\big) \ket{4}_R
\end{align}    
\end{subequations}

\paragraph{Local reference phase errors.--} Consider an error that corresponds to a measurement of the number of atoms in the reference modes, $\sum_i r^\dagger_i r_i$. This collapses the logical state for instance into the states
\begin{subequations}
\begin{align}
\label{eq:example-states}
    \ket{1,0}_L&\rightarrow \ket{111} (\ket{110}+\ket{101}+\ket{011})\ket{0}_R,\\
\ket{0,1}_L&\rightarrow (\ket{110}+\ket{101}+\ket{011})\ket{111}\ket{0}_R.
\end{align}    
\end{subequations}
That is, an initial superposition of logical basis states
\begin{align}
    \ket{\Psi} = \alpha \ket{1,0}_L + \beta \ket{0,1}_L\, ,
\end{align}
collapses into, e.g.
\begin{align}
    \ket{\Psi} \rightarrow &[\alpha \ket{111} (\ket{110}+\ket{101}+\ket{011})\nonumber \\ &+ \beta (\ket{110}+\ket{101}+\ket{011})\ket{111}]\ket{0_R}\, .
\end{align}
This shows that the logical quantum information is preserved. This holds analogously for all measurement outcomes. Re-encoding the state with stabilizer measurements gives back the original logical quantum state. Hence, this code fulfills the error correction conditions for local phase errors on the reference modes, which can be removed by measuring the total atom number in the reference, such that the phase error is promoted to a global phase. 

\paragraph{Number-conserving logical gates.--} Furthermore, this example also illustrates, that operations that conserve the number of logical fermions, can be implemented without operating on the reference modes. For instance, we consider here the example of a logical $f$\textsc{swap}, which maps the two logical states \eqref{eq:example-states} onto each other. This can be achieved by physically swapping all system modes of the first block with their counterpart in the second block. This operation therefore does not involve operating on the reference. The more general case is then discussed below.

\subsubsection{Number-conserving logical gates}

\paragraph{Logical gates without reference operations.--}As we illustrated for a simple two-mode example above, any logical operation in $\mathcal{H}^C_{N_L}$ can be done without operating on the reference. This is also true in the general multi-mode case. For all gates acting with density operators on a logical mode, this follows from the fact that their logical annihilation and creation operators appear in pairs, i.e.
\begin{align}
    C^\dagger_b C_{b} 
    & =[c_1c_2c_3^\dagger+ c_1c_2^\dagger c_3+c_1^\dagger c_2 c_3 + c_1^\dagger c_2^\dagger c_3^\dagger]_b\nonumber\\& [c_3c_2c_1 + c_3^\dagger c_2^\dagger c_1 + c_3^\dagger c_2 c_1^\dagger + c_3 c_2^\dagger c_1^\dagger ]_b\nonumber\\
&= (1-n_{b,1})(1-n_{b,2})n_{b,3} + (1-n_{b,1})n_{b,2}(1-n_{b,3})\nonumber\\
& + n_{b,1}(1-n_{b,2})(1-n_{b,3})+n_{b,1}n_{b,2}n_{b,3}\nonumber\\
&= 4n_{b,1}n_{b,2}n_{b,3} +n_{b,1}+n_{b,2}+n_{b,3}\nonumber\\
& -2n_{b,1}n_{b,3}-2n_{b,2}n_{b,3}-2n_{b,2}n_{b,1}\, .
\end{align}
With $n_{b,i}=c_{b,i}^\dag c_{b,i}=s_{b,i}^\dag s_{b,i}$ we see that this operator only involves system modes. 
Furthermore, the logical tunneling gate can be written as
\begin{align}
     e^{i\theta(C^\dagger_b C_{b'} + h.c.)} = \begin{cases}
         &\cos(\theta)+ i\sin(\theta) [f\textsc{swap}]_{L,bb'} \\
         &\qquad \text{if}\, N_b+N_{b'} =1\\ &1 \quad \text{else}\, .
     \end{cases}
\end{align}
Here $N_b=C_b^\dag C_b$.
As shown below, $[f\textsc{swap}]_L$ does not involve $R$, such that we can realize the logical tunneling gate without using the reference.

\subsubsection{Implementation of logical gates} In this subsection we discuss the implementations of the logical operations in terms of physical operations. We discuss transversal gates as well as non-transversal gates which we implement with additional qubit ancillas.

\paragraph{Transversal logical gates.--} We demonstrate the $f$\textsc{swap} gate to be transversal.
Consider its action on two physical fermionic modes
\begin{subequations}
\begin{align}
    \ket{00}&\overset{f\textsc{swap}}{\longrightarrow}\ \ket{00}\\
    \ket{01}&\overset{f\textsc{swap}}{\longrightarrow} \ket{10}\\
    \ket{10}&\overset{f\textsc{swap}}{\longrightarrow} \ket{01}\\
    \ket{11}&\overset{f\textsc{swap}}{\longrightarrow} - \ket{11}\, .
\end{align}
\end{subequations}
Its action can also be described as
\begin{align}
    (f\textsc{swap})_{ij} \,c_i\, (f\textsc{swap})_{ij} = c_j\, ,
\end{align}
and analogously for $c^\dagger_i$ as well as the mode $j$. Defining fermionic exchange across code blocks $b$ and $b'$, we get
\begin{align}
    (f\textsc{swap})_{bb',i}\, c_{b,i}\, (f\textsc{swap})_{bb',i} = c_{b',i}\, ,
\end{align}
and subsequently
\begin{align}
    &[\prod_i (f\textsc{swap})_{bb',i}]\, c_{b,0}c_{b,1}c_{b,2}\, [\prod_i (f\textsc{swap})_{bb',i}]\nonumber\\
    &= c_{b',0}c_{b',1}c_{b',2}\, .
\end{align}
The same reasoning applies to any combination of fermion modes within block $b$; they simultaneously get exchanged with the corresponding modes of block $b'$. Hence, the same holds true for logical operators, i.e.
\begin{align}
    [\prod_i (f\textsc{swap})_{bb',i}]\, C_{b}\, [\prod_i (f\textsc{swap})_{bb',i}]
    = C_{b'}\, ,
\end{align}
and therefore, the logical exchange is transversal
\begin{align}
    f\textsc{swap}_{L,bb'} = \prod_i (f\textsc{swap})_{bb',i}\, .
\end{align}
The $f$\textsc{swap} of the physical modes $(f\textsc{swap})_{bb',i}$ are physically implementable with a classical exchange of the corresponding tweezers, they do not involve operations on the reference modes. Therefore the logical $f$\textsc{swap} does not involve operations on the reference modes.

Consider similarly the tunneling with angle~$\pi/2$. It relates to $f$\textsc{swap} as follows
\begin{align}
    e^{i\frac{\pi}{2}(s_i^\dagger s_j + \text{H.c.})} =  (f\textsc{swap})_{ij} \mathrm{S}_i \mathrm{S}_j\, ,
\end{align}
where $\mathrm{S}_i = \exp(i\tfrac{\pi}{2}s^\dagger_i s_i)$ is the fermionic version of the $\mathrm{S}$-gate. Since $\mathrm{S}$-gates are not transversal in repetition codes, neither are $\pi/2$-tunneling gates. However, $\mathrm{S}$ can be made transversal, e.g. in fermion codes based on 2D color codes, which renders the tunneling transversal too.

\paragraph{Non-transversal logical gates.--}
To implement non-transversal logical gates we employ stable ancilla qubits, that is, we assume that the required operations can be performed perfectly on the ancilla qubits. The following operations entangle the qubit with the fermionic system such that the gate gets transferred to the logical fermions. Specifically, consider the first circuit shown in \Fig{fig:gates}($b$). Using a qubit language and considering a single logical fermionic mode here, i.e. $\mathrm{Z}_L = \exp(i\pi C^\dagger C)$, the first element is $\mathrm{H}_a (\mathrm{C}_a\mathrm{Z}_L)\mathrm{H}_a$, which results in 
\begin{align}
\ket{\psi_L}\ket{0}_a&\overset{\mathrm{H}_a}{\longrightarrow}  \ket{\psi_L}\frac{\ket{0}_a+\ket{1}_a}{\sqrt{2}}\nonumber\\ &\overset{\mathrm{C}_a\mathrm{Z}_L }{\longrightarrow} \frac{\ket{\psi_L}\ket{0}_a+\mathrm{Z}_L\ket{\psi_L}\ket{1}_a}{\sqrt{2}}\nonumber\\ &\overset{\mathrm{H}_a }{\longrightarrow} \frac{1+\mathrm{Z}_L}{2}\ket{\psi_L}\ket{0}_a + \frac{1-\mathrm{Z}_L}{2}\ket{\psi_L}\ket{1}_a\, ,
\end{align}
i.e. the gates entangle the qubit state with the logical-$\mathrm{Z}_L$ state of the block. Hence, operations which are performed on the qubit now, simultaneously operate on the logical state of the logical fermionic mode. The second application of $\mathrm{H}_a (\mathrm{C}_a\mathrm{Z}_L) \mathrm{H}_a$ undoes the entanglement of system and ancilla and the qubit gate is transferred onto the logical block. Specifically,
\begin{align}
&\mathrm{T}\left[\frac{1+\mathrm{Z}_L}{2}\ket{\psi_L}\ket{0}_a + \frac{1-\mathrm{Z}_L}{2}\ket{\psi_L}\ket{1}_a\right]\nonumber\\
&=\frac{1+\mathrm{Z}_L}{2}\ket{\psi_L}\ket{0}_a + \frac{1-\mathrm{Z}_L}{2}\ket{\psi_L}e^{i\frac{\pi}{4}}\ket{1}_a\nonumber\\
&\overset{\mathrm{H}_a }{\longrightarrow} \frac{1}{\sqrt{2}}\left[\frac{1+\mathrm{Z}_L}{2}+\frac{1-\mathrm{Z}_L}{2}e^{i\frac{\pi}{4}}\right]\ket{\psi_L}\ket{0}_a \nonumber\\
&\quad + \frac{1}{\sqrt{2}}\left[\frac{1+\mathrm{Z}_L}{2}-\frac{1-\mathrm{Z}_L}{2}e^{i\frac{\pi}{4}}\right]\ket{\psi_L}\ket{1}_a \nonumber\\
&\overset{\mathrm{C}_a\mathrm{Z}_L }{\longrightarrow} \frac{1}{\sqrt{2}}\left[\frac{1+\mathrm{Z}_L}{2}+\frac{1-\mathrm{Z}_L}{2}e^{i\frac{\pi}{4}}\right]\ket{\psi_L}\ket{0}_a \nonumber\\
&\quad + \frac{1}{\sqrt{2}}\left[\frac{1+\mathrm{Z}_L}{2}+\frac{1-\mathrm{Z}_L}{2}e^{i\frac{\pi}{4}}\right]\ket{\psi_L}\ket{1}_a\nonumber\\
&= \left[\frac{e^{i\frac{\pi}{4}}+1}{2}+\frac{1-e^{i\frac{\pi}{4}}}{2}\mathrm{Z}_L\right]\ket{\psi_L}\frac{\ket{0}_a+\ket{1}_a}{\sqrt{2}} \nonumber\\
&\overset{\mathrm{H}_a }{\longrightarrow} \mathrm{T}_L\ket{\psi_L}\ket{0}_a \, ,
\end{align}
where $\mathrm{T}_L = \exp(i\tfrac{\pi}{4}C^\dagger C)$ is the logical fermionic version of the $\mathrm{T}$-gate on a single logical mode~$b$. The same reasoning can be applied to the remaining circuits shown in~\Fig{fig:gates}.

\subsubsection{Reference error correction}
As stated in the main text, and illustrated on the example of two logical modes above, local phase errors on the reference modes can also be corrected if we consider $M_L$ logical modes with fixed logical fermion number. It follows from the discussion above that any logical basis state with a fixed logical fermion number $N_L$ can be obtained from any logical initial state $\ket{\psi_\mathrm{i}}$ in $\mathcal{H}_{N_L}^C$ by applications of unitaries $U_s$ which involve system modes $s_{b,i}$ only. Suppose, we choose $\ket{\psi_\mathrm{i}} = \ket{0,..,0,1,..,1}_L$, with $N_L$ logical fermions, and $U_s^\mathbf{j}$ is a product of $f$\textsc{swap} gates which bring us to any logical basis state, i.e. $\ket{\mathbf{j}}_L = U_s^\mathbf{j} \ket{\psi_\mathrm{i}}$. Here, $\mathbf{j} = j_1,\dots,j_{M_L}$ labels the logical fermion configuration. 
Notably, all basis states are orthonormal, i.e. $\langle \mathbf{j}|\mathbf{i}\rangle_L = \delta_{\mathbf{j}\mathbf{i}}$ as two different Fock states $\ket{\mathbf{i}}_L\neq \ket{\mathbf{j}}_J$ have orthogonal occupations of the physical modes $s_i$. 

We consider single local phase errors $E_{Ra} = 1-2r^\dagger_ar_a$ in the reference, which commute with the logical parity operator $P_b =1-2C^\dagger_b C_b$ for all blocks $b$. Hence, the basis remains orthogonal under application of the error term
\begin{align}
    \bra{\mathbf{j}}_L E^\dagger_{Ra} E_{Rb}  \ket{\mathbf{i}}_L = \delta_{\mathbf{i}\mathbf{j}}\bra{\mathbf{j}}_L E^\dagger_{Ra} E_{Rb}  \ket{\mathbf{j}}_L\, .
\end{align}
This can be seen explicitly as follows. Suppose $\ket{\mathbf{i}}_L$ and $\ket{\mathbf{j}}_L$ with $\mathbf{i}\neq \mathbf{j}$ differ in the parity of the $b$-th block, e.g. $P_b\ket{\mathbf{i}}_L = \ket{\mathbf{i}}_L$ and $P_b\ket{\mathbf{j}}_L = -\ket{\mathbf{j}}_L$. Then, we get 
\begin{align}
   \bra{\mathbf{j}}_L E^\dagger_{Ra} E_{Rb}  \ket{\mathbf{i}}_L &=   \bra{\mathbf{i}}_L P_b E^\dagger_{Ra} E_{Rb}  \ket{\mathbf{j}}_L\nonumber\\
   &= \bra{\mathbf{j}}_L  E^\dagger_{Ra} E_{Rb}  P_b\ket{\mathbf{j}}_L\nonumber\\
   &= - \bra{\mathbf{j}}_L  E^\dagger_{Ra} E_{Rb}  \ket{\mathbf{j}}_L\nonumber\\
   & = 0\, ,
\end{align}
for $\mathbf{i}\neq \mathbf{j}$. Then, since $U^\mathbf{j}_s$ involves system modes only, they commute with the reference errors,
\begin{align}
    [U^\mathbf{j}_s,E_{Ra}] = 0\, .
\end{align}
Therefore, we obtain
\begin{align}
    \bra{\mathbf{j}}_L E^\dagger_{Ra} E_{Rb}  \ket{\mathbf{i}}_L &= \delta_{\mathbf{i}\mathbf{j}}\bra{\psi_\mathrm{i}}(U^\mathbf{j}_s)^\dagger E^\dagger_{Ra} E_{Rb} U^\mathbf{j}_s \ket{\psi_\mathrm{i}} \nonumber\\ &=\delta_{\mathbf{i}\mathbf{j}}\bra{\psi_\mathrm{i}} E^\dagger_{Ra} E_{Rb}  \ket{\psi_\mathrm{i}} \nonumber\\
    &=\delta_{\mathbf{i}\mathbf{j}}\mathcal{C}_{ab} ,
\end{align}
which is the Knill-Laflamme error correction condition.

\subsection{Measuring the fermionic exchange phase}
In this section, we show the analytical calculation of the circuit in \Fig{fig:experiment} directly on the level of logical fermions without any errors. The circuit performs the following operations: first the ancilla is rotated as
\begin{align}
    \ket{1,1,0}_L\ket{0}_a &\rightarrow \ket{1,1,0}_L\frac{\ket{0}_a+\ket{1}_a}{\sqrt{2}} 
\end{align}
After that, the operations are controlled on $\ket{1}_a$, i.e.
\begin{align}
    \ket{1,1,0}_L\ket{1}_a
    &\rightarrow -i\ket{0,1,1}_L\ket{1}_a \nonumber\\&\rightarrow -i^2\ket{1,0,1}_L \ket{1}_a\nonumber\\&\rightarrow -i^3\ket{1,1,0}_L \ket{1}_a\, .
\end{align}
Thus the final state is given by
\begin{align}
    \ket{1,1,0}_L\frac{\ket{0}_a-i^3\ket{1}_a}{\sqrt{2}} = \ket{1,1,0}_L\ket{y}_a,
\end{align}
Measuring the ancilla in the $y$-basis, yields $+1$, while a corresponding bosonic particle would yield $-1$. In general, for any exchange phase $\Theta$, one gets $\ket{\psi}\propto \ket{0}-i\exp(i\Theta)\ket{1}$, and hence
\begin{align}
    \langle Y\rangle_a = - \cos(\Theta)\, .
\end{align}
Measurement of the ancilla therefore reveals the particle exchange statistics. A phase error on one of the modes during the circuit may change the ancilla state and therefore corrupt the signal. 
\clearpage
\end{document}